\documentclass[prb, aps, twocolumn, superscriptaddress]{revtex4}
\usepackage{graphicx}
\usepackage{bm}
\usepackage{amssymb}
\usepackage{amsmath}
\usepackage{amsfonts}
\usepackage{color}

\newcommand {\ket} [1] {| #1 \rangle}
\newcommand {\bkt} [1] {\langle #1 \rangle}

\newcommand {\pd} [2] {\frac{\partial #1}{\partial #2}}
\newcommand {\td} [2] {\frac{d #1}{d #2}}

\newcommand {\ra} {\rightarrow}

\newcommand* {\vek}[1]{{\bm{#1}}}

\usepackage [english]{babel}

\begin{document}
\title{Anomalous spin precession and spin Hall effect in semiconductor quantum wells}
\author{Xintao Bi}
\affiliation{ICQD, Hefei National Laboratory for Physical Sciences at the Microscale, University of Science and Technology of China, Hefei, Anhui 230026, China}
\author{Peiru He}
\affiliation{ICQD, Hefei National Laboratory for Physical Sciences at the Microscale, University of Science and Technology of China, Hefei, Anhui 230026, China}
\author{E.~M.~Hankiewicz}
\affiliation{Institut f\"ur Theoretische Physik und Astrophysik, Universit\"at W\"urzburg, 97074 W\"urzburg, Germany}
\author{R. Winkler}
\affiliation{Materials Science Division, Argonne National Laboratory, Argonne, IL 60439.}
\affiliation{Northern Illinois University, De Kalb, IL 60115.}
\author{Giovanni Vignale}
\affiliation{Department of Physics and Astronomy, University of Missouri, Columbia, Missouri 65211}
\author{Dimitrie Culcer}
\affiliation{ICQD, Hefei National Laboratory for Physical Sciences at the Microscale, University of Science and Technology of China, Hefei, Anhui 230026, China}
\affiliation{School of Physics, The University of New South Wales, Sydney 2052, Australia}

\begin{abstract}
  SO (SO) interactions give a spin-dependent correction
  $\hat{\bm r}_{so}$ to the position operator, referred to as the
  \textit{anomalous} position operator. We study the contributions
  of $\hat{\bm r}_{so}$ to the spin-Hall effect (SHE) in quasi
  two-dimensional (2D) semiconductor quantum wells with strong band
  structure SO interactions that cause spin precession. The skew
  scattering and side-jump \textit{scattering} terms in the SHE
  vanish, but we identify two additional terms in the SHE, due to
  $\hat{\bm r}_{so}$, which have not been considered in the
  literature so far. One term reflects the modification of spin
  precession due to the action of the external electric field (the
  field drives the current in the quantum well), which produces, via
  $\hat{\bm r}_{so}$, an effective magnetic field perpendicular to
  the plane of the quantum well. The other term reflects a similar
  modification of spin precession due to the action of the
  electric field created by random impurities, and appears in a
  careful formulation of the Born approximation. We refer to these
  two effects collectively as \textit{anomalous spin precession} and
  we note that they contribute to the SHE to the first order in the
  SO coupling constant even though they formally appear to be of
  second order. In electron systems with weak momentum scattering,
  the contribution of the anomalous spin precession due to the
  external electric field equals 1/2 the usual side-jump SHE, while
  the additional impurity-dependent contribution depends on the form
  of the band structure SO coupling. For band structure SO coupling linear in
  wave vector the two anomalous spin precession contributions
  cancel. For band structure SO coupling cubic in wave vector,
  however, they do not cancel, and the anomalous spin precession
  contribution to the SHE can be detected in a high-mobility 2DEG
  with strong SO coupling. In 2D hole systems both anomalous
  spin precession contributions vanish identically.
\end{abstract}
\date{\today}

\maketitle

\section{Introduction}

In systems with strong spin-orbit (SO) interactions an electric
field generates a transverse spin-current\cite{Dyakonov71, Hirsch99,
Zhang00, Murakami03, Sinova04, Handbook, WinRev,
Culcer_SteadyState_PRB07, Culcer_Generation_PRL07, TsePRB05,
Malshukov_Edge_PRL05, GalitskiPRB06}: this phenomenon is referred to
as the spin-Hall effect (SHE). For the past ten years, the SHE has been
a source of new ideas for magneto-electronic
devices\cite{Ralph_MgnSwc_SHE_11} aimed at integrating semiconductor
and magnetic technologies, facilitating efficient information
processing and quantum computing
architectures.\cite{Wu_SpinDyn_PhysRep10, Fabian_Slov07,
Awschalom07, Zutic04} These visions have stimulated a large volume
of experimental and theoretical work.\cite{Gu_Renorm_PRL10,
Tanaka_NJP09, Guo_IntrinSH_PRL08, Guigou_Interf_PRB11,
Cheng_SkewScat_JPCM08, Dugaev_PRB10, Gradhand_Tail_PRB10,
Gradhand_ExtriSH_PRL10, Dyakonov_SpinAcc_MagRes_PRL07,
Hiroshi_PRL09, Kovalev_AHE_PRB09, Liu_PRB06, Loss_Diffus_PRB10,
Rashba_Semi08, Schwab_2DEG_EPL10, Niu_sidejump_PRB11}
Experimentally, the SHE was initially studied in semiconductors,
\cite{Kato04, Sih05, Wunderlich05, Stern06, Stern08} but has since
expanded to novel materials such as HgTe-based quantum wells,
\cite{Bruene_Ballistic_NatPhys10} topological insulators and
graphene, and $d$-band metals.\cite{Gao_TI_GiantSHE_PRL11,
Dora_TI_Gfn_Hall_PRB11, Culcer_TI_PhysE12} It is often simpler to
measure the \textit{inverse} spin-Hall
effect,\cite{Saitoh_InverSH_APL06} where a spin current generates a
transverse charge current, which is detected by conventional means.
The inverse SHE has been observed in Al, \cite{Valenzuela_Nat06} Pt
wires at room temperature, \cite{Takahashi_Revese_PRL07} hybrid
FePt/Au devices, \cite{Takahashi_GSH_NatMater08} Au films with Pt
impurities, \cite{Gu_Surf_PRL10} permalloy/normal metal bilayers,
\cite{Mosendz_PRB10} GaAs multiple quantum wells,\cite{Werake_PRL11}
and Cu with Ir impurities. \cite{Niimi_PRL11} For a review of recent
experimental work on the SHE in Pt see
Ref.~\onlinecite{Ralph_Pt_SHE_Rvw_11}. Observation of the inverse
SHE has recently been reported even in a weakly SO coupled
material such as Si. \cite{Saitoh_InverSH_arX11}

SO coupling may be present in the band structure and in the impurity
potentials. Band-structure SO interactions become important in
structures lacking a center of inversion when SO interactions lift
spin degeneracy. \footnote{Interestingly, however, the strongest
spin Hall effect to date is found in inversion-symmetric metals.} If
the underlying crystal lattice lacks a center of inversion the
material is said to possess bulk inversion asymmetry (BIA). In
low-dimensionsional systems the confinement potential can be made
asymmetric, in which case one speaks of structure inversion
asymmetry (SIA). In this paper we consider exclusively quasi
two-dimensional semiconductor systems that lack a center of
inversion due to BIA \footnote{For example, a zincblende structure
gives rise to Dresselhaus SO coupling, which we take as a
prototypical SO coupling where needed.} and/or SIA giving rise to
Rashba SO coupling. In these systems the band structure SO
interaction is represented by a Hamiltonian $H = (\hbar/2) \,
\bm{\sigma}\cdot\bm{\Omega}_{\bm k}$ describing the interaction of
the spin with an effective wave vector-dependent magnetic field
$\bm{\Omega}_{\bm k}$. This can be
$\vek{\Omega}_\vek{k}^\mathrm{BIA}$ or
$\vek{\Omega}_\vek{k}^\mathrm{SIA}$. The spin precesses about this
field with frequency $\Omega_{\bm k} \equiv |\bm{\Omega}_{\bm k}|$.
Different physical regimes are distinguished by the value of the
product of $\Omega_{\bm k}$ with the momentum relaxation time
$\tau_p$. In the ballistic regime (clean limit) $\Omega_{\bm k}
\tau_p \rightarrow \infty$. The weak momentum scattering regime is
characterized by $\Omega_{\bm k}\tau_p \gg 1$, while in the strong
momentum scattering regime $\Omega_{\bm k}\tau_p \ll 1$.

SO interactions arise, quite generally, from a spin-dependent
correction $\hat{\bm r}_{so}$ to the position operator,
\cite{Nozieres73, Foldy50} whose general form is
\begin{equation}
  \hat{\bm r}_{so} = \lambda \, \bm{\sigma}\times \bm{\omega}_{\bm k},
\end{equation}
where $\lambda$ is a material-specific parameter, ${\bm \sigma}$ is
the vector of Pauli spin matrices, and ${\bm \omega}_{\bm k}$ takes
different forms for different systems, as well as for electrons and
holes in the same system.\footnote{The formalism presented in this
paper applies to quasi-2D systems. No matter whether we have
electrons or holes, in 2D they can be projected on a $2\times 2$
subspace, though in general the two spinor components in this
Hilbert space cannot be interpeted as plain ``spin up'' or ``spin
down''. Instead these spinor components denote some entangled
spin-orbital motion.} We note that $\vek{\Omega}_\vek{k}$ and
$\vek{\omega}_\vek{k}$ are not independent of each other. The
$\vek{\omega}_\vek{k}$ entering the corrected position operator is
inherently related to $\vek{\Omega}_\vek{k}^\mathrm{SIA}$
characterizing the Rashba SO coupling through a term of the form
$\vek{\Omega}_\vek{k}^\mathrm{SIA} = \lambda \vek{\omega}_\vek{k}
\times \nabla V$. The presence of $\hat{\bm r}_{so}$ results in
corrections to the interaction between charge carriers and electric
fields, which include impurities and external electric fields. Thus,
in addition to the band structure SO interaction, one must take into
account SO interactions arising from the external electric field and
from the electron-impurity potential. The interplay between these
interactions in the SHE is quite a complicated subject. It has
received a lot of attention in recent years, yet, as we will see, it
is not yet completely understood.

Perhaps the most intuitive mechanism of SHE is the one known as
\textit{skew scattering}, i.e., the asymmetric scattering of up and
down spins by impurities. \cite{Kohn57, TsePRL06, Engel05} Next, we
have the so-called \textit{side-jump scattering} term, \cite{Luttinger,
Berger70a, Berger72, Nozieres73} which consists of two equal terms,
one reflecting the correction to the band energy due to the
spin-dependent interaction with the electric field, the other
reflecting the renormalization of the carrier trajectory during
collisions. Diagrammatic formulations naturally recover the two
side-jump scattering terms through the vertex renormalization of
spin and charge currents, as Ref.~\onlinecite{TsePRL06}
demonstrated. An analytical derivation of the side jump from the
Kubo formula was presented in Ref.~\onlinecite{Hankiewicz_QW_PRL06}.
Furthermore, Ref.\ \onlinecite{TsePRB05} identified skew-scattering
and side-jump scattering within a drift-diffusion approach. More
recently, side jump scattering was derived starting from the quantum
Liouville equation for the single-particle spin density-matrix.
\cite{Culcer_SideJump_PRB10}

The analysis of the SHE becomes considerably more complicated when
\textit{both} band-structure and impurity-potential induced SO
interactions are present. This problem was first addressed by Tse
and Das Sarma, \cite{TsePRB06} who employed the diagrammatic Kubo
formula and considered band structure SO coupling of the linear
Rashba form. They found that the skew scattering contribution to the
SHE vanished for arbitrarily small value of the band structure SO
coupling, while a term equal to half the usual side-jump scattering
SHE survived. \footnote{Refs.~\onlinecite{Hu06, Wu08} also found a
vanishing skew scattering contribution to the SHE, and similar
conclusions were reached concerning the vanishing of the skew
scattering contribution to the 2D \textit{anomalous} Hall effect.
\cite{Sinitsyn_SkewScat_PRL07}} This is in contrast to the result
obtained in Ref.~\onlinecite{Hankiewicz08} that both the side-jump
and the skew scattering contributions vanish for arbitrarily small
values of the band structure SO coupling, as long as
impurity-induced (Elliot-Yafet) spin relaxation is neglected. These
two results are reconciled by taking into account the SO
contribution to the electron-impurity self-energy
diagram,~\cite{Raimondi_AnnPhys12} which recovers the vanishing of
the side-jump and skew scattering contributions found in
Ref.~\onlinecite{Hankiewicz08}.

The principal question identified in Ref.~\onlinecite{TsePRB06} was
the paradox of the non-analyticity of the spin Hall conductivity,
which appears to change discontinuously as soon as the band
structure SO coupling is turned on. This paradox was finally solved
in Ref.~\onlinecite{Raimondi09} by the introduction of an
impurity-induced (Elliott-Yafet) spin relaxation rate
$1/\tau_\mathrm{EY}$, which led to a spin Hall conductivity of the
form
\begin{equation}\label{SHE-Crossover}
  \sigma^z_{yx}=
  \frac{[\sigma^z_{yx}]_{ss}+[\sigma^z_{yx}]_{sj}}
  {1+\tau_\mathrm{EY}/\tau_\mathrm{DP}}
\end{equation}
where $\tau_\mathrm{DP}$ is the Dyakonov-Perel relaxation time
associated with the band structure SO coupling and given by
$\tau_\mathrm{DP}^{-1}=\langle\Omega_k^2 \rangle \tau_p$, where
$\tau_p$ is the momentum relaxation time and the angular bracket
denotes an average over the momentum distribution. The above formula
exhibits a smooth crossover between the sum of skew-scattering (ss)
and side-jump scattering (sj) contributions, when the band structure spin
precession $\Omega_k$ is neglected, and zero when $\Omega_k\tau_p
\gg 1$, i.e. when the band-structure SO interaction is much stronger
than the electron-impurity interaction (see also
Ref.~\onlinecite{Raimondi_AnnPhys12}).

However, this is not the end of the story. The work described above
was limited to band-structure SO couplings that are linear in wave
vector ${\bm k}$. The aim of this work is to provide a consistent
framework for treating band structure and impurity SO effects in
quasi two-dimensional quantum wells for {\it any form of the band
structure SO interaction} in the weak momentum scattering regime
$\Omega_{\bm k}\tau_p \gg 1$. To this end, we construct a rigorous
theory of the interplay of spin precession due to band structure SO
coupling and SO coupling due to impurities. We start from the
quantum Liouville equation and derive a kinetic equation for the
spin density matrix, which captures the effects of band-structure
spin precession and $\hat{\bm r}_{so}$ on an equal footing. We focus
from the very beginning on the weak momentum scattering regime
$\Omega_{\bm k} \tau_p \gg 1$. Under this assumption, we do not have
to worry about the finite Elliot-Yafet scattering rate that appears
in Eq.~(\ref{SHE-Crossover}): we are in the regime $\tau_\mathrm{EY}
\gg \tau_\mathrm{DP}$. But, while Eq.~(\ref{SHE-Crossover})
predicts, in this limit, a vanishing spin Hall conductivity for
linear-in-${\bm k}$ band-structure SO interaction, we will show that
a finite spin Hall conductivity can survive for different forms of
that SO interaction.

More precisely, we find that, in the weak momentum scattering
regime, skew scattering and side jump scattering still give zero
SHE. At the same time, we identify two additional contributions to
the SHE stemming from $\hat{\bm r}_{so}$. These contributions have
been overlooked in the literature thus far. One contribution arises
from the impurity potential, and is found in the Born approximation
when scattering terms of second order in SO are taken into account.
This contribution can be viewed as a modification of the band
structure precession frequency due to the electron-impurity interaction.
The second contribution is scattering-\textit{independent}. Its
origin lies in the spin-dependent interaction with the external
electric field brought about by $\hat{\bm r}_{so}$. This has the
form of an interaction between each carrier and an effective
magnetic field. The
carrier spin precesses in this effective magnetic field in such a
way that an out-of-plane spin component is generated, which
contributes to the SHE. We refer to these two effects collectively
as \textit{anomalous spin precession}. The impurity-induced
anomalous spin precession term gives an out-of plane component of
the effective magnetic field. This is precisely what distinguishes
anomalous spin precession from the usual side-jump scattering
term, which vanishes in the presence of spin-precession. Remarkably,
these effects contribute to the SHE in the first order in the SO
coupling constant even though they formally appear to be of second
order. The external electric field part of the anomalous spin
precession term appears to be universal in electronic systems in the
clean limit.

In electron systems with band structure SO linear in ${\bm k}$ the
sum of the two anomalous spin precession terms vanishes. In hole
systems both additional terms are zero independently. Nevertheless,
the anomalous spin precession term in the SHE in general survives, and we demonstrate its existence explicitly
in 2D electron systems with band structure SO described by the cubic
Dresselhaus model. In this model we find the total SHE conductivity in the
clean limit to be [see Eq.\ (\ref{CubicDresSHE}) below]
\begin{equation}
    \sigma^z_{yx} \approx - \frac{e}{16\pi} +
    \frac{n_ee\lambda}{4}.
\end{equation}
The term $\propto \lambda$ is linear in the
electron number density, while the band-structure SO contribution in
the weak momentum scattering regime is density-independent. The cubic Dresselhaus
SO interaction term is strong in a wide electron quantum well at high density $n_e$. 
Although the experimental situation is more complicated than the above
formula suggests (see Sec.\ \ref{sec:expt}), and involves the non-trivial
interplay of linear and cubic SO terms, we find that in a high-mobility 2D electron gas based on InSb, anomalous
spin precession accounts for most of the spin-Hall conductivity. Our
results are therefore relevant to experiments and help to distinguish
different contributions to the SHE.

Contributions to the SHE purely from band structure SO are well known. \cite{Handbook} We do not discuss them explicitly here, except in the practical case of experimental observation (Sec.\ \ref{sec:expt}). The focus of this work is on the contributions to the SHE due to $\hat{\bm
r}_{so}$, and the central result is that, aside from the well-known
skew scattering and side-jump scattering terms, two additional
contributions -- the anomalous spin precession terms -- are present
when band structure SO is nonzero. To our knowledge, this is the
first work that proves that $\hat{\bm r}_{so}$ can give rise to a
spin-Hall current through a mechanism unrelated to scattering. We
work up to third order in the impurity potential, and, in order to
recover all contributions, we consider terms of second order in the SO
coupling. Our results are valid in the weak momentum scattering
limit, yet in the Appendix we prove rigorously that a
non-analyticity in the strong momentum scattering limit is cured by
introducing the Elliott-Yafet spin relaxation time
$\tau_\mathrm{EY}$, as was done in Ref.\
\onlinecite{Raimondi_AnnPhys12}.

The outline of this paper is as follows. In Sec.~\ref{sec:BandHam}
we present the band Hamiltonian and in Sec.~\ref{sec:r} we discuss
the effective position operator. In Sec.~\ref{sec:kineq} we derive
the general form of the kinetic equation starting with the quantum
Liouville equation, and discuss the various scattering terms. In
Sec.~\ref{sec:DM} we discuss the non-equilibrium correction to the
density matrix, demonstrating that a new, scattering-independent
driving term due to $\hat{\bm r}_{so}$ is present. The general
solution to the kinetic equation is presented in Sec.\
\ref{sec:sol}, demonstrating that the skew scattering and side-jump
scattering terms give zero contributions to the SHE. All SHE
contributions due to $\hat{\bm r}_{so}$ are listed for commonly
employed models of SO coupling. An explanation of anomalous spin
precession is given in Sec.~\ref{sec:disc}, which is followed by a
detailed discussion of the experimental situation in
Sec.~\ref{sec:expt}, and the summary and conclusions.

\section{Band Hamiltonian}
\label{sec:BandHam}

In the crystal-momentum representation, the band Hamiltonian
$\hat{H}_0$ in the effective mass approximation has the general form
\begin{equation}
     H_{0{\bm k}} = H_\mathrm{kin} + H_{so}
    \equiv H_\mathrm{kin} + \frac{\hbar}{2} \, {\bm \sigma} \cdot {\bm
      \Omega}_{\bm k},
\end{equation}
for an arbitrary SO interaction. The kinetic energy term
$H_\mathrm{kin} = \varepsilon_{0{\bm k}} \openone \equiv
\frac{\hbar^2k^2}{2m^*} \, \openone$, where $\openone$ is the
identity matrix in spin space and $m^*$ the carrier effective mass.
The spin-dependent term in the Hamiltonian $H_{so}$ is treated as a
perturbation with respect to the kinetic energy term. The
eigen-energies are written as $\varepsilon_{{\bm k}\pm} =
\varepsilon_{0{\bm k}} \pm (\hbar \Omega_{\bm k}/2)$.

For quasi-2D systems we may have different contributions to SO
coupling that are relevant in different regimes. \cite{Winkler2003}
For 2D spin-1/2 electron systems with SIA, the band structure
contains the linear Rashba Hamiltonian
\begin{equation}
  H_{R1} = \alpha_1 \, (\sigma_x k_y - \sigma_y k_x) = \alpha_1
  i(k_- \sigma_+ - k_+ \sigma_-),
\end{equation}
where $k_\pm \equiv k_x \pm ik_y$ and $\sigma_\pm \equiv (\sigma_x
\pm i\sigma_y)/2$. For the most common case of a (001) surface BIA
has two contributions, the linear Dresselhaus term
\begin{equation}
  H_{D1} = \beta_1 (\sigma_y k_y - \sigma_x k_x) = - \beta_1 (k_+
  \sigma_+ + k_- \sigma_-),
\end{equation}
and the cubic Dresselhaus term
\begin{equation}
  \arraycolsep 0.3ex
  \begin{array}[b]{rl}
    H_{D3} & = \beta_3 (\sigma_x k_x
    k_y^2 - \sigma_y k_y k_x^2) \\[1ex] & = \beta_3
    [k_-(k_+^2-k_-^2) \sigma_+ + k_+(k_-^2-k_+^2) \sigma_-].
  \end{array}
\end{equation}
In a quantum well with well width $w$ we have approximately $\beta_1
= \beta_3 (\pi/w)^2$ (Ref.~\onlinecite{Winkler2003}). This implies
that the linear Dresselhaus term often dominates in more narrow
electron systems with smaller density (i.e., small Fermi wave
vector), whereas the cubic Dresselhaus term may dominate in wider
quantum wells with a larger density. Experiments can be designed to
focus on these different regimes. Even in the latter case we
typically remain in the electric quantum limit, where only the
lowest subband of the quantized motion in $z$ direction is occupied. \cite{Wunderlich05} In the following we
  will focus on this regime.

For 2D heavy-hole systems SO coupling due to SIA is dominated by the
cubic Rashba Hamiltonian,
\begin{equation}
  \arraycolsep 0.3ex
  \begin{array}[b]{rl}
    H_{R3} & = \alpha_3 [k_y (k_y^2-3k_x^2) \sigma_x + k_x (k_x^2 -
      k_y^2) \sigma_y] \\[1ex] & = \alpha_1 i(k_+^3 \sigma_- - k_-^3
    \sigma_+).
  \end{array}
\end{equation}
BIA in 2D heavy-hole systems on a (001) surface contains the
$k$-linear term
\begin{equation}
  H_{D1'} = \gamma_1 (\sigma_x k_x + \sigma_y k_y) = \gamma_1 (k_+
  \sigma_- + k_- \sigma_+),
\end{equation}
and the cubic Dresselhaus term
\begin{equation}
  \arraycolsep 0.3ex
  \begin{array}[b]{rl}
     H_{D3'} & =  \gamma_3 (k_x^2 + k_y^2)
    (\sigma_x k_x + \sigma_y k_y) \\[1ex] & = \gamma_3 (k_+^2 k_-
    \sigma_- + k_-^2 k_+ \sigma_+).
  \end{array}
\end{equation}
For the terms cubic in $k$, we restricted ourselves to the dominant
contributions due to SIA and BIA. $H_{D1'}$ and $H_{D3'}$ are often
comparable in magnitude.

\section{Effective position operator}
\label{sec:r}

The SO interaction appears when transforming from the Dirac to the
Pauli equation by means of the Foldy-Wouthuysen transformation.
\cite{Foldy50} Under this transformation, the position operator in
spin-1/2 systems becomes
\begin{equation}
  \hat{\bm r}_{phys} = \hat{\bm r} + \hat{\bm r}_{so},
\end{equation}
where the SO part $ \hat{\bm r}_{so}$ is expressed in terms of the
vector ${\bm \sigma}$ of Pauli spin matrices. We refer to $\hat{\bm
  r}_{so}$ as the anomalous position operator.

The general form for the anomalous position operator, valid for both
2D electron and 2D hole systems, is
\begin{equation}\label{rso}
  \hat{\bm r}_{so} = \lambda \, {\bm \sigma}\times{\bm \omega}_{{\bm
      k}},
\end{equation}
where $\lambda$ and $\bm{\omega}_{\bm k}$ are different for
electrons and holes. For 2D electrons ${\bm \omega}_{\bm k} = {\bm
  k}$, and
\begin{equation}\label{rsoe}
  \hat{\bm r}_{{so}} = \lambda_1 \, {\bm \sigma}\times{\bm k},
\end{equation}
assuming $\lambda_1 k_F^2 \ll 1$. For 2D hole systems the
correction to the position operator has the form
\begin{equation}\label{rsoh}
  \hat{\bm r}_{so} = \lambda_3 \, {\bm \sigma}\times{\bm
    \omega}_{{\bm k}3},
\end{equation}
where ${\bm \omega}_{{\bm k}3} = k^3 \, (\cos 3 \theta, \sin 3
\theta, 0)$, assuming $\lambda_3 k_F^6 \ll 1$.

Consider a general scalar potential $V (\hat{\bm r})$. Under the
Foldy-Wouthuysen transformation it transforms to $V (\hat{\bm
  r}_{phys})$, which, to first order in $\hat{\bm r}_{so}$, takes
the form
\begin{equation}
  V (\hat{\bm r}_{phys}) = V (\hat{\bm{r}}) + \frac{1}{2} \,
  [\bm{\nabla} V(\hat{\bm{r}}) \cdot \hat{\bm r}_{so} + \hat{\bm
      r}_{so}\cdot \nabla V(\hat{\bm{r}})].
\end{equation}
Therefore, as a result of this transformation, both the potential
due to an applied electric field and the impurity scattering
potential acquire spin-dependent terms.

Let $U({\bm r})$ denote the scattering potential, which represents
elastic scattering off charged impurities and static defects (but
not phonons or electrons)
\begin{equation}
   U({\bm r}) = \sum_I \bar{U} ({\bm r} - {\bm R}_I),
\end{equation}
where ${\bm R}_I$ indexes the random locations of the impurities and
the scattering potential due to a single impurity is denoted by
$\bar{U}({\bm r})$. In Fourier space, the matrix elements of $U({\bm
  r})$ are
\begin{equation}
   U_{{\bm k}{\bm k}'} = \bar{U}_{{\bm k}{\bm k}'}
  \sum_I e^{i({\bm k} - {\bm k}')\cdot{\bm R}_I}.
\end{equation}
and the potential due to a single impurity is written as
\begin{equation}
     \bar{U}_{{\bm k}{\bm k}'} =  
    \mathcal{U}_{{\bm k}{\bm k}'} \openone + \mathcal{V}_{{\bm
        k}{\bm k}'},
\end{equation}
where $\mathcal{U}_{{\bm k}{\bm k}'}$ represents the matrix element
of the potential due to a single impurity between plane waves, while
$\mathcal{V}_{{\bm k}{\bm k}'}$ is the spin-dependent part arising
from $\hat{\bm r}_{so}$. Both have units of energy $\times$ volume.
The strength of the disorder potential is characterized by the
impurity density $n_i$. The matrix elements of the spin-dependent
part of the impurity potential in reciprocal space are
\begin{equation}
     \mathcal{V}_{{\bm k}{\bm k}'} =  -
    \frac{i\lambda}{2} \, {\bm \sigma} \cdot ({\bm \omega}_{\bm k}
    \times{\bm k}' - {\bm \omega}_{{\bm k}'} \times{\bm k}) \,
    \mathcal{U}_{{\bm k}{\bm k}'}.
\end{equation}
In 2D the spin dependent term in $\mathcal{V}_{{\bm k}{\bm k}'}$
points out of the plane for both electron and hole systems.

Interaction with a static, uniform external electric field ${\bm E}$
is contained in
\begin{equation}\label{HE}
\arraycolsep 0.3ex
\begin{array}{rl}
\displaystyle H_{E{\bm k}{\bm k}'} = & \displaystyle (e{\bm E}\cdot\hat{\bm r})_{{\bm k}{\bm k}'} \openone + e \, ({\bm E}\cdot\hat{\bm r}_{so})_{{\bm k}{\bm k}} \delta_{{\bm k}{\bm k}'} \\ [1ex]
= & \displaystyle ie{\bm E}\cdot\pd{}{{\bm k}} \, \delta({\bm k} - {\bm k}') \, \openone + \frac{1}{2} \, {\bm \sigma}\cdot{\bm \Delta}_{{\bm k}} \delta_{{\bm k}{\bm k}'}.
\end{array}
\end{equation}
with $\openone$ the identity matrix in spin space, and ${\bm
\Delta}_{{\bm k}}$ arises from the anomalous position operator.
\cite{Sinitsyn_AHE_Review_JPCM08} From Eq.\ (\ref{rso}),
\begin{equation}\label{Delta_el}
\arraycolsep 0.3ex
\begin{array}{rl}
\displaystyle {\bm \Delta}_{{\bm k}} = & \displaystyle 2 e \lambda \,  {\bm \omega}_{\bm k} \times {\bm E}.
\end{array}
\end{equation}
It follows from the preceding discussion that ${\bm \Delta}_{{\bm
k}}$ has different forms in electron and hole systems.

The anomalous position operator accounts for impurity SO coupling
and for band structure SO coupling due to SIA. To see the latter,
consider the SO coupling due to the full potential $V_{tot}$ acting
on the system. In a 2D system we can divide $V_{tot} = V_{ext} +
V_{QW} + U$, where $V_{ext}$ is the applied electric field, $V_{QW}$
the $z$-direction confinement, and $U$ the impurity potential
introduced above. The total potential $V_{tot}$ gives rise to SO
coupling, which in reciprocal space is contained in
\begin{equation}
  H_{so, {\bm k}} = \lambda_n {\bm \sigma} \cdot {\bm k} \times {\bm
    \nabla} (V_{ext} + U) + \lambda_n {\bm \sigma} \cdot {\bm k}
  \times \hat{\bm z} \, \bigg( \pd{V_{QW}}{z} \bigg).
\end{equation}
In the second term we can incorporate the average $\bkt{\partial
V_{QW}/\partial{z}}$ over the quantum well into an effective SO
constant $\alpha$, giving the Rashba SO coupling. \footnote{This
simplified discussion has neglected the complex relationship between
the quantum well average of ${\bm \nabla}V_{QW}$ and Ehrenfest's
theorem, which is covered in detail in
Ref.~\onlinecite{Winkler2003}.} This clarifies the relationship
between $\alpha$ and $\lambda$ and shows that, knowing the form of
the Rashba Hamiltonian in a certain system, one can deduce the form
of $\hat{\bm r}_{so}$ in that system.

The full Hamiltonian is $H^{tot}_{{\bm k}} = H_{0{\bm k}} + H_{E{\bm
k}{\bm k}'} + U_{{\bm k}{\bm k}'}$. The spin current operator
$\hat{j}^i_j$ corresponding to spin component $i$ flowing in the
direction $j$ is
\begin{equation}\label{jij}
  \hat{j}^i_j = \frac{\hbar^2 k_j}{2m} \, \sigma_i.
\end{equation}
In addition to the contribution from the band Hamiltonian, the
velocity operator has two additional terms, discussed in detail in
Ref.~\onlinecite{Culcer_SideJump_PRB10}. The first stems from the
spin-dependent interaction with the external electric field $H_{E
{\bm k}\lambda}$, while the second arises from the spin-dependent
term $\mathcal{V}_{{\bm k}{\bm k}'}$ in the impurity potential.
These two cancel, as they represent the net force acting on the
system. \cite{Culcer_SideJump_PRB10} They will not be explicitly
considered in what follows.

\section{Kinetic equation}
\label{sec:kineq}

The formalism presented here parallels that originally formulated in
Refs.~\onlinecite{Culcer_SteadyState_PRB07,
Culcer_Generation_PRL07}. The Liouville equation for the density
operator $\hat\rho$ is projected onto the basis $\{ \ket{{\bm k}}
\}$, with $\rho_{{\bm k}{\bm k}'} = f_{{\bm k}} \, \delta_{{\bm
k}{\bm k}'} + g_{{\bm k}{\bm k}'}$, where $g_{{\bm k}{\bm k}'}$ is
off-diagonal in ${\bm k}$, and \textit{all} quantities are matrices
in spin space. These satisfy
\begin{subequations}
  \begin{eqnarray}
    \label{eq:FermiJfk}
    \td{f_{\bm k}}{t} + \frac{i}{\hbar} \, [\hat{H}_0,
    \hat{f}]_{{\bm k}{\bm k}} &=& - \frac{i}{\hbar} \, [\hat{U},
    \hat{g}]_{{\bm k}{\bm k}} \\ [1ex] \label{eq:Jfk} \td{g_{{\bm
    k}{\bm k}'}}{t} + \frac{i}{\hbar} \, [\hat{H}_0, \hat{g}]_{{\bm
    k}{\bm k}'} &=& - \frac{i}{\hbar} \, [\hat{U}, \hat{f}]_{{\bm
    k}{\bm k}'} - \frac{i}{\hbar} \, [\hat{U}, \hat{g}]_{{\bm k}{\bm
    k}'}, \hspace*{2em}
  \end{eqnarray}
\end{subequations}
We focus on variations which are slow on the scale of the momentum
relaxation time, and solve for $g_{{\bm k}{\bm k}'}$ as an expansion
in the impurity potential, which can be performed to any desired
order. Very generally $f_{\bm k}$ satisfies
\begin{equation}\label{kineq}
  \td{f_{\bm k}}{t} + \frac{i}{\hbar} \, [\hat{H}_0, \hat{f}]_{{\bm
      k}{\bm k}} + \hat{J}(f_{\bm k}) = 0.
\end{equation}
The total scattering term $\hat{J} (f_{\bm k}) =
\hat{J}_\mathrm{Born}(f_{\bm k}) + \hat{J}_{ss}(f_{\bm k})$, where
in the first Born approximation
\begin{equation}
  \label{JBorn}
  \hat{J}_\mathrm{Born}(f_{\bm k}) =
  \frac{1}{\hbar^2}\bigg\langle\int_0^{\infty} \!\! dt' \, [\hat U, e^{-
      i \hat{H}_0 t'/\hbar}[\hat U, \hat f]\, e^{ i
        \hat{H}_0 t'/\hbar}] \bigg\rangle_{{\bm k}{\bm k}},
\end{equation}
and $\bkt{\ldots}$ represents averaging over impurity
configurations. In the second Born approximation we obtain the
additional skew scattering term
\begin{widetext}
\begin{equation}
  \label{Jss3rd}
  \hat{J}_{ss}(f_{\bm k}) = - \frac{i}{\hbar^3}
  \bigg\langle\int_0^{\infty} dt' \int_0^{\infty} dt'' [\hat U, e^{-
      i \hat{H}_0 t'/\hbar}[\hat U, e^{- i \hat{H}_0
          t''/\hbar}[\hat U, \hat f]\, e^{i \hat{H}_0
          t''/\hbar} ]\, e^{i \hat{H}_0
        t'/\hbar}]\bigg\rangle_{{\bm k}{\bm k}}.
\end{equation}
We expand $\hat{J}_\mathrm{Born} (f_{\bm k})$ in $\Omega_{\bm k}$
and $\lambda$. We retain the leading term plus terms to first order
in $\Omega_{\bm k}$, first order in $\lambda$, and the second-order
term in $\Omega_{\bm k} \lambda$. Thus $\hat{J}_\mathrm{Born}
(f_{\bm k})$ can be written as a perturbation expansion in
$\Omega_{\bm k}$ and $\lambda$ in the form
\begin{equation}
  \hat{J}_\mathrm{Born}(f_{\bm k}) = \hat{J}_0(f_{\bm k}) +
  \hat{J}_\mathrm{\Omega}(f_{\bm k}) + \hat{J}_\mathrm{sj}(f_{\bm
    k}) + \hat{J}_\mathrm{\Omega\lambda}(f_{\bm k}).
\end{equation}
The leading term in $\hat{J}_\mathrm{Born} (f_{\bm k})$ is the
scalar $\hat{J}_0(f_{\bm k})$, which is the customary
Born-approximation scattering term appearing in the Boltzmann
equation. It is found by taking Eq.\ (\ref{JBorn}) and considering
only the scalar parts of $\hat{H}_0$ (i.e. $H_\mathrm{kin}$) and
$\hat{U}$ (i.e. $\mathcal{U}_{{\bm k}{\bm k}'}$), and in 2D takes
the form
\begin{equation}
  \arraycolsep 0.3ex
  \begin{array}{rl}
    \displaystyle \hat{J}_0 (f_{{\bm k}}) = & \displaystyle
    \frac{n_im^*}{\hbar^3} \, \int\frac{d\theta'}{2\pi}\,
    |\mathcal{U}_{{\bm k}{\bm k}'}|^2 (f_{{\bm k}} - f_{{\bm k}'}).
  \end{array}
\end{equation}
Next, we have the term in $\hat{J}_\mathrm{Born} (f_{\bm k})$ to
first order in $\Omega_{\bm k}$ (i.e. due to band-structure SO
coupling), which is found by considering the spin-dependent part of
$\hat{H}_0$ and the scalar part of $\hat{U}$. It gives rise to a
well-known scattering term, referred to here as
$\hat{J}_\mathrm{\Omega} (f_{{\bm k}})$.
\cite{Culcer_SteadyState_PRB07, Culcer_Generation_PRL07} We only
require its action on the scalar part of the density matrix, $n_{\bm
k}$, given by
\begin{equation}
  \arraycolsep 0.3ex
  \begin{array}{rl}
    \displaystyle \hat{J}_\Omega \, (n_{\bm k}) = & \displaystyle
    \frac{\pi}{\hbar} \int \frac{d^2k'}{(2\pi)^2} \,
    |\mathcal{U}_{{\bm k}{\bm k}'}|^2 \,(n_{{\bm k}} - n_{{\bm
        k}'})\, {\bm \sigma}\cdot({\bm \Omega}_{{\bm k}} - {\bm
      \Omega}_{{\bm k}'}) \, \pd{}{\varepsilon_0} \,
    \delta(\varepsilon_{0{\bm k}} - \varepsilon_{0{\bm k}'}) .
  \end{array}
\end{equation}
This term is relevant only in determining the band-structure SO
contribution to the spin current, which has been studied previously,
and is not pertinent to the discussion presented in this work and
will not be given. Following on, in the side-jump scattering term
$\hat{J}_{sj}(f_{\bm k})$ we take the scalar part of $\hat{H}_0$ and
the spin-dependent part of $\hat{U}$. The electric field ${\bm E}$
is also finite in this term: without it $\hat{J}_{sj}(f_{\bm k})$
would vanish. \footnote{This corresponds to the well-known argument
that there is no skew scattering in the Born approximation
\cite{Landau3}.} Because ${\bm E}$ is nonzero,
$\hat{J}_{sj}(f_{\bm k})$ acts on the equilibrium density matrix
$f_{0{\bm k}}$. It has two parts, which have been determined in
Ref.~\onlinecite{Culcer_SideJump_PRB10}. We use the notation of
Ref.~\onlinecite{Culcer_SideJump_PRB10}. We write
$\hat{J}_\mathrm{sj} \, (n_{\bm k}) = \hat{J}^{a}_\mathrm{sj} \,
(n_{\bm k}) + \hat{J}^{b}_\mathrm{sj} \, (n_{\bm k})$. The first
part of the side-jump scattering term, referred to as
$\hat{J}^{a}_\mathrm{sj} \, (n_{\bm k})$, arises from the change in
the band energy due to the spin-dependent energy of interaction with
${\bm E}$
\begin{equation}\label{Jsja}
  \arraycolsep 0.3ex
  \begin{array}{rl}
    \displaystyle \hat{J}^{a}_\mathrm{sj} \, (n_{\bm k}) = &
    \displaystyle \frac{2\pi n_i}{\hbar}\,
    \int\!\frac{d^2k'}{(2\pi)^2} \: |\mathcal{U}_{{\bm k}{\bm
        k}'}|^2(n_{\bm k} - n_{{\bm k}'})\, \frac{1}{2} \,
    \bm{\sigma}\cdot(\bm{\Delta}_{{\bm k}} - \bm{\Delta}_{{\bm k}'})
    \, \pd{}{\varepsilon_{0{\bm k}}} \, \delta(\varepsilon_{0{\bm
        k}} - \varepsilon_{0{\bm k}'}).
  \end{array}
\end{equation}
The second part, $\hat{J}^b_\mathrm{sj} \, (n_{\bm k})$, reflects
the spin-dependent change in the carrier position during collisions
\begin{equation}\label{Jsjb}
  \arraycolsep 0.3ex
  \begin{array}{rl}
    \displaystyle \hat{J}^{b}_\mathrm{sj} \, (n_{\bm k}) = &
    \displaystyle \frac{i n_i \pi e\bm{E}}{\hbar} \cdot
    \int\!\frac{d^2k'}{(2\pi)^2} \mathcal{U}_{{\bm k}{\bm k}'}
    \bigg(\pd{\mathcal{V}_{{\bm k}'{\bm k}}}{{\bm k}'} +
    \pd{\mathcal{V}_{{\bm k}'{\bm k}}}{{\bm k}} \bigg) \, (n_{{\bm
        k}} - n_{{\bm k}'}) \, \pd{}{\varepsilon_{0{\bm k}'}} \,
    \delta(\varepsilon_{0{\bm k}} - \varepsilon_{0{\bm k}'}) + h.c.
  \end{array}
\end{equation}
Both parts of the side jump scattering term are $\propto \sigma_z$.

The scattering term $\hat{J}_{\Omega \lambda}(n_{\bm k})$ reads
\begin{equation}
  \hat{J}_{\Omega \lambda}(n_{\bm k}) = \frac{ \pi n_i }{\hbar }
  \!\! \int \!\! \frac{d^dk'}{(2\pi)^d} \, [\,\bm{\sigma} \cdot {\bm
      \Omega}_{\bm k'} \, , \, \mathcal{V}_{{\bm k}{\bm k}'}] \,
  \mathcal{U}_{{\bm k}{\bm k}'} \,( n_{\bm k} - n_{\bm k'}) \,
  \frac{\partial}{\partial\varepsilon_{0{\bm k}}}
  \delta(\varepsilon_{0{\bm k}} - \varepsilon_{0{\bm k}'}).
\end{equation}
\end{widetext}
The physical meaning of this term is as follows. During a scattering
process, an incoming spin has a well-defined spin direction, given
by ${\bm \Omega}_{\bm k}$, which represents the band-structure SO
coupling at wave vector ${\bm k}$. Because the scattering potential
is also spin dependent, the incoming spin is rotated during
scattering by an amount that is proportional to $\mathcal{V}_{{\bm
k}{\bm k}'}$, the impurity SO coupling. This scattering term
therefore represents spin rotations during collisions induced by the
impurity SO coupling, the rotation being evident from its commutator
structure.

Even though we are doing perturbation theory to first order in the
SO interaction terms $\lambda$ and $\Omega_{\bm k}$, spin precession
makes it necessary to include driving terms to order
$\lambda\Omega_{\bm k}$, since these terms also yield contributions
to the spin current $\propto \lambda$ only, i.e. to first order in
the impurity SO coupling. The necessity of including terms $\propto
\lambda\Omega_{\bm k}$ will become apparent when we discuss
explicitly the solution for $S_{E{\bm k}}$ introduced below, during
which it will emerge that spin precession introduces a factor of
$1/\Omega_{\bm k}$.

Beyond the first Born approximation we retain the leading term
$\hat{J}_{ss}(f_{\bm k})$, in which $\lambda$ is finite but the
electric field ${\bm E} = 0$, which is customarily responsible for
skew scattering. \cite{Sinitsyn_AHE_Review_JPCM08} To first order in
$\lambda$, the real part of this term reduces to
\begin{widetext}
\begin{equation}
  \arraycolsep 0.3ex
  \begin{array}{rl}
    \displaystyle \hat{J}_{ss}(n_{\bm k}) = & \displaystyle -
    \frac{3\pi^2 n_i
      \lambda}{\hbar}\int\frac{d^2k'}{(2\pi)^2}\int\frac{d^2k''}{(2\pi)^2}
    \, \mathcal{U}_{{\bm k}{\bm k}'} \mathcal{U}_{{\bm k}'{\bm
        k}''}\mathcal{U}_{{\bm k}''{\bm k}} \, {\bm \sigma} \cdot
    ({\bm \omega}_{{\bm k}} \times{\bm k}' - {\bm \omega}_{{\bm k}'}
    \times{\bm k}) \, (n_{{\bm k}'} - n_{{\bm k}''}) \delta
    (\varepsilon_{0{\bm k}} - \varepsilon_{0{\bm k}''}) \delta
    (\varepsilon_{0{\bm k}} - \varepsilon_{0{\bm k}'}).
  \end{array}
\end{equation}
In 2D systems, in which both ${\bm k}$ and ${\bm \omega}_{{\bm k}}$
are in the $xy$-plane, the skew scattering term is $\propto
\sigma_z$.

\section{Non-equilibrium density matrix}
\label{sec:DM}

In a constant uniform electric field ${\bm E}$ the density matrix is
$f_{\bm{k}} = f_{0 \bm{k}} + f_{E {\bm{k}}} $. The equilibrium
density matrix is given by
\begin{equation}
 f_{0{\bm k}} = {\textstyle\frac{1}{2}} \,
[f_\mathrm{FD}(\varepsilon_{{\bm k}+}) + 
f_\mathrm{FD}(\varepsilon_{{\bm k}-})] +
{\textstyle\frac{1}{2}} \, 
[f_\mathrm{FD}(\varepsilon_{{\bm k}+}) - f_\mathrm{FD}(\varepsilon_{{\bm
      k}-})] \, {\bm \sigma} \cdot \hat{\bm \Omega}_{\bm k}, 
\end{equation}
with $f_\mathrm{FD}$ the Fermi-Dirac distribution function, while
$f_{E {\bm{k}}}$ is due to ${\bm E}$. To first order in ${\bm E}$
the correction $f_{E {\bm{k}}}$ satisfies
\begin{equation}\label{eq:boltze}
  \pd{f_{E {\bm{k}}}}{t} + \frac{i}{\hbar}\, [H_{{\bm k}}, f_{E
  {\bm{k}}}] + \hat{J}\, (f_{E {\bm{k}}}) = \frac{e{\bm
  E}}{\hbar}\cdot\pd{f_{0{\bm k}}}{{\bm k}} - \frac{i}{2\hbar}\,
  [{\bm \sigma}\cdot{\bm \Delta}_{{\bm k}}, f_{0{\bm k}}] .
\end{equation}
\end{widetext}
The term $(e{\bm E}/\hbar)\cdot(\partial f_{0{\bm k}}/\partial{\bm
k})$ corresponds to the usual streaming term in the Boltzmann
equation. The second term on the RHS of Eq.\ (\ref{eq:boltze})
appears due to the anomalous position operator and is $\propto
\lambda$.

We write $f_{\bm k} = n_{\bm k}\openone + S_{\bm k}$, where $S_{\bm
k}$ is a $2 \times 2$ Hermitian matrix, and correspondingly $f_{E
{\bm{k}}} = n_{E {\bm{k}}} \openone + S_{E {\bm{k}}}$ and $f_{0{\bm
k}} = n_{0{\bm k}}\openone + S_{0{\bm k}}$. The expectation values
of the spin current operator is found from $S_{E {\bm{k}}}$. The
term $(e{\bm E}/\hbar)\cdot(\partial f_{0{\bm k}}/\partial{\bm k})$
may be decomposed into a scalar part $(e{\bm E}/\hbar)\cdot(\partial
n_{0{\bm k}}/\partial{\bm k})$ and a spin-dependent part $(e{\bm
E}/\hbar)\cdot(\partial S_{0{\bm k}}/\partial{\bm k})$. The
spin-dependent part has been studied previously,
\cite{Culcer_SteadyState_PRB07, Culcer_Generation_PRL07} and is
responsible for current-induced spin polarizations and spin currents
arising from the band-structure SO coupling. It will not be
discussed in this work.

The non-equilibrium correction to the scalar part of the density
matrix, $n_{E {\bm{k}}}$, is determined from
\begin{equation}\label{eq:nE}
  \pd{n_{E{\bm k}}}{t} + \hat J_0 \, (n_{E {\bm{k}}}) = \frac{e{\bm
      E}}{\hbar}\cdot\pd{n_{0{\bm k}}}{{\bm k}}.
\end{equation}
The solution to this equation is well known, and reads $n_{E{\bm k}}
= (e{\bm E}\tau_p/\hbar)\cdot (\partial n_{0{\bm k}}/\partial{\bm
k})$, with $\tau_p$ the momentum relaxation time. Once this solution
is found, the spin-dependent scattering terms $\hat{J}_{ss}$,
$\hat{J}_{sj}$ and $\hat{J}_{\Omega\lambda}$ act on $n_{E{\bm k}}$
and produce additional effective driving terms for $S_{E{\bm k}}$.
(The method used is the same as in Ref.~\onlinecite{Culcer_TI_AHE_PRB11}.)

We seek the solution for $S_{E{\bm k}}$ to first order in $\lambda$
which we denote by $S_{E{\bm k}\lambda}$. Specifically, including
the contribution due to ${\bm \Delta}_{\bm k}$ from
Eq.\ (\ref{eq:boltze}), it is found from
\begin{widetext}
\begin{equation}
  \arraycolsep 0.3ex
  \begin{array}{rl}
    \displaystyle \pd{S_{E {\bm k}\lambda}}{t} + & \displaystyle
    \frac{i}{\hbar}\, [H_{{\bm k}}, S_{E {\bm k}\lambda}] +
    \hat{J}_0 \, (S_{E {\bm k}\lambda}) = - \hat{J}_{ss}(n_{E{\bm
        k}}) - \hat{J}_{sj}(n_{E{\bm k}}) -
    \hat{J}_{\Omega\lambda}(n_{E{\bm k}}) - \frac{i}{\hbar}\,
        [H_{E{\bm k}\lambda}, S_{0{\bm k}}]. 
  \end{array}
\end{equation}
\end{widetext}
We specialize to short-range impurities henceforth, without loss of
generality. The potential of a single impurity in Fourier space is
written as
\begin{subequations}
  \begin{eqnarray}
    \displaystyle \bar{U}_{{\bm k}{\bm k}'} & = & \displaystyle
    \mathcal{U} \openone + \mathcal{V}_{{\bm k}{\bm k}'} \\ [1ex]
    \displaystyle \mathcal{V}_{{\bm k}{\bm k}'}& = & \displaystyle -
    \frac{i\lambda \mathcal{U}}{2} \, {\bm \sigma} \cdot ({\bm
    \omega}_{{\bm k}} \times{\bm k}' - {\bm \omega}_{{\bm k}'}
    \times{\bm k}),
  \end{eqnarray}
\end{subequations}
where the Fourier transform $\mathcal{U}_{{\bm k}{\bm k}'}$ has
become the constant $\mathcal{U}$. We write $\hat{J}_0 (f_{{\bm k}})
= (f_{{\bm k}} - \overline{f_{{\bm k}}})/\tau$, with the overline
denoting an angular average over the directions of $\hat{\bm k}$,
which in 2D indicates an average over the polar angle $\theta$,
\begin{equation}
  \overline{X} \equiv \int\frac{d\theta}{2\pi} \, X,
\end{equation}
and the momentum relaxation time $\tau_p \equiv \tau$, given by
\begin{equation}\label{tau}
  \arraycolsep 0.3ex
  \begin{array}{rl}
    \displaystyle \frac{1}{\tau} = & \displaystyle
    \frac{n_im^*\mathcal{U}^2}{\hbar^3}.
  \end{array}
\end{equation}

We discuss the driving terms in more detail. Firstly,
\begin{equation}\label{Jssomega}
  \arraycolsep 0.3ex
  \begin{array}{rl}
    \displaystyle - \hat{J}_{ss}(n_{E{\bm k}}) = & \displaystyle
    \frac{3n_i \lambda m^{*2} |\mathcal{U}|^3}{4\hbar^5} {\bm
      \sigma} \cdot \overline{({\bm \omega}_{{\bm k}} \times{\bm k}'
      - {\bm \omega}_{{\bm k}'} \times{\bm k}) \, n_{E{\bm k}'}},
  \end{array}
\end{equation}
where the overline denotes averaging over $\theta'$ and the
integration over $k'$ forces $k' = k$. We have established that this
term is $\propto \sigma_z$, and inspection of Eq.\ (\ref{Jssomega})
reveals that this term is an odd function of ${\bm k}$.

The anomalous interaction with ${\bm E}$ gives rise to \textit{two}
driving terms. The first arises from the side-jump scattering term,
which was determined in Ref.~\onlinecite{Culcer_SideJump_PRB10}. For
both electrons and holes this takes the form
\begin{equation}\label{Jsjomega}
  \arraycolsep 0.3ex
  \begin{array}{rl}
    \displaystyle - \hat{J}_{sj}(n_{E{\bm k}}) = & \displaystyle -
    \frac{1}{\tau} \, \bm{\sigma}\cdot\bm{\Delta}_{{\bm k}}
    \,\delta(\varepsilon_{0{\bm k}} - \varepsilon_F).
  \end{array}
\end{equation}
This term is also odd in ${\bm k}$. \footnote{For 2D hole systems
  and short-range impurities this term still contains the factor of
  2. In the general case of long-range scattering it is not certain
  that the result can be simplified in this way.} An additional
driving term comes from the commutator of $\frac{1}{2} \, {\bm
  \sigma}\cdot{\bm \Delta}_{{\bm k}}$ with the density matrix. Given
that ${\bm \Delta}_{{\bm k}}$ is already first-order in ${\bm E}$ we
require only the equilibrium density matrix $f_{0{\bm k}}$. We
expand $f_{0{\bm k}} = f_\mathrm{FD}(\varepsilon_{{\bm k}}) \openone +
(\hbar/2) \, {\bm \sigma} \cdot {\bm \Omega}_{\bm k} \,
\pd{f_\mathrm{FD}(\varepsilon_{{\bm k}})}{\varepsilon_{{\bm k}}}$, where
the first term is a scalar, and at temperature $T = 0$ we can write
\begin{equation}\label{dsj}
  - \frac{i}{\hbar}\, [H_{E{\bm k}}^{sj}, f_{0{\bm k}}] =
  \frac{1}{2}\, \delta(\varepsilon_{0{\bm k}} - \varepsilon_F) \,
  {\bm \sigma}\cdot{\bm \Omega}_{\bm k} \times {\bm \Delta}_{{\bm
      k}}.
\end{equation}
Notice that this term is zero in the absence of spin precession,
when $f_{0{\bm k}}$ is a scalar and the commutator vanishes.

The remaining driving term is $- \hat{J}_{\Omega\lambda}(n_{E{\bm
    k}})$. For 2D electron systems,
\begin{widetext}
\begin{equation}
  - \hat{J}_{\Omega\lambda}(n_{E{\bm k}}) = \frac{i \pi \lambda n_i
    |\mathcal{U}|^2}{\hbar} \!\! \int \!\! \frac{d^2k'}{(2\pi)^2} \,
  [\,\bm{\sigma} \cdot {\bm \Omega}_{\bm k'} \, , {\bm
      \sigma}\cdot{\bm k}\times{\bm k}' ] \, ( n_{E{\bm k}} -
  n_{E{\bm k}'}) \, \frac{\partial}{\partial\varepsilon_{0{\bm k}}}
  \delta(\varepsilon_{0{\bm k}} - \varepsilon_{0{\bm k}'}).
\end{equation}
For 2D hole systems,
\begin{equation}
  - \hat{J}_{\Omega\lambda}(n_{E{\bm k}}) = \frac{i\lambda \pi n_i
    |\mathcal{U}|^2}{2\hbar } \!\! \int \!\! \frac{d^2k'}{(2\pi)^2}
  \, [\,\bm{\sigma} \cdot {\bm \Omega}_{\bm k'}, {\bm \sigma} \cdot
    ({\bm \omega}_{{\bm k} 3} \times{\bm k}' - {\bm \omega}_{{\bm
        k}' 3} \times{\bm k})] \,(n_{E{\bm k}} - n_{E{\bm k}'}) \,
  \frac{\partial}{\partial\varepsilon_{0{\bm k}}}
  \delta(\varepsilon_{0{\bm k}} - \varepsilon_{0{\bm k}'}).
\end{equation}
\end{widetext}

\section{Solution of the kinetic equation}
\label{sec:sol}

We summarize first the general solution to the kinetic equation for
short range impurities and weak momentum scattering. We denote the
driving terms generically by $\mathcal{D}_{E{\bm k}\lambda}$ in this
section. Let the component $i$ of the spin operator be denoted by
$\hat{s}_i = (\hbar/2) \, \sigma_i$. The spin density is ${\rm Tr
}\, \rho \hat s_i = {\rm Tr }\, \bar\rho \hat s_i $, where the
overbar denotes an angular average as above, thus $\bar\rho$ is the
isotropic part of the density matrix. Similarly, the spin current
operator $\hat{j}^i_j$ has been defined in Eq.\ (\ref{jij}). Because
it is odd in ${\bm k}$ its expectation value yields ${\rm Tr }\,
\rho \hat{j}^i_j = {\rm Tr }\, (\rho - \bar\rho) \hat{j}^i_j$.
Consequently, the isotropic part of the spin density matrix
determines the spin density, while the anisotropic part of the
density matrix determines the spin current. It is therefore
convenient to divide the spin density matrix into $S_{E{\bm
k}\lambda} = \overline{S_{E{\bm k}\lambda}} + T_{E{\bm k}\lambda}$,
the isotropic part being $\overline{S_{E{\bm k}\lambda}}$ (which
gives the spin density) and the anisotropic part $T_{E{\bm
k}\lambda}$ (which gives the spin current). From the quantum
Liouville equation, we obtain a set of coupled equations for
$\overline{S_{E{\bm k}\lambda}}$ and $T_{E{\bm k}\lambda}$ for
short-range impurities, which are solved rigorously in Appendix
\ref{sec:ST}. Here we just quote the solution. Letting
$\displaystyle \mathcal{D}_{E{\bm k}\lambda} = \frac{1}{2} \, {\bm
\sigma} \cdot {\bm d}_{E{\bm k}\lambda}$, we find for $\Omega_{\bm
k}\tau \gg 1$
\begin{equation}\label{TlargeSO}
  T_{E{\bm k}\lambda}= \frac{1}{2} \, {\bm \sigma} \cdot
  \bigg(\frac{\hat{\bm \Omega}_{\bm k}}{\Omega_{\bm k}}\bigg) \times
       [{\bm d}_{E{\bm k}\lambda} +
         \overline{\mathcal{A}}^{-1}(\overline{{\bm d}_{E{\bm
               k}\lambda}} - \overline{\mathcal{A}{\bm d}_{E{\bm
               k}\lambda}})],
\end{equation}
where the (dimensionless) matrix $\mathcal{A}$ is given by
$\mathcal{A}_{ij} = (\delta_{ij} - \hat{\Omega}_i\hat{\Omega}_j)$,
and $T_{E{\bm k}\lambda}$ as found in Eq.\ (\ref{TlargeSO}) gives
the spin current in the weak momentum scattering limit. Finally, we
take the electric field ${\bm E} \parallel \hat{\bm x}$, the
spin-Hall conductivity is defined by $j^z_y = \sigma^z_{yx}E_x$, and
we abbreviate the spin-Hall conductivity due to $T_{E{\bm
k}\lambda}$ simply by $\sigma_\lambda$.

The appearance of the $\Omega_k$ in the denominator of
Eq.\ (\ref{TlargeSO}) is a crucial feature of this solution. It
demonstrates the need to retain scattering terms $\propto \lambda
\Omega_{\bm k}$ that are formally of second order in the SO
coupling.

\subsection{Skew scattering and side-jump scattering}

We recall that, as shown in Eqs.\ (\ref{Jssomega}) and
(\ref{Jsjomega}), both $\hat{J}_\mathrm{ss} \, (n_{E{\bm k}})$ and
$\hat{J}_\mathrm{sj} \, (n_{E{\bm k}})$ are odd in ${\bm k}$.
Therefore the driving terms due to $\hat{J}_\mathrm{ss} \, (n_{E{\bm
k}})$ and $\hat{J}_\mathrm{sj} \, (n_{E{\bm k}})$ yield corrections
to $S_{E{\bm k}\lambda}$ that are even in ${\bm k}$. Since the spin
current operator $\hat{j}^i_j$ is odd in ${\bm k}$, simple power
counting in Eq.\ (\ref{TlargeSO}) reveals that $\hat{J}_\mathrm{ss}
\, (n_{E{\bm k}})$ and $\hat{J}_\mathrm{sj} \, (n_{E{\bm k}})$ do
not give a spin current in the weak momentum scattering regime. We
can develop a physical understanding of this fact. In the absence of
spin precession, skew scattering and side-jump scattering separate
up-spins from down-spins. When band structure SO interactions are
present, each spin precesses about an \textit{effective} magnetic
field which depends on ${\bm k}$, thus it is not conserved.
Electrons are driven by the external field and collide with
impurities, with up-spins scattering predominantly in one direction
and down-spins predominantly in the other direction. The spins then
travel towards the edges of the sample, yet they are subjected to
the action of the band structure SO effective field, which causes
them to precess. Upon arriving at the edge the spins are completely
randomized. Therefore, very generally, side-jump scattering and skew
scattering do not give rise to a spin current in 2D systems.

\begin{table}[tbp]
  \caption{\label{tab:disentangle} $\hat{\bm r}_{so}$ contributions
    to the SHE in units of $n_ee\lambda$ for $\Omega\tau \gg 1$.
    Here $e^-$ ($h^+$) stands for electrons (holes), while
    ``band SO'' abbreviates ``band-structure SO''.}
    $\arraycolsep 1em
  \begin{array}{c@{\hspace{2em}}cccc} \hline\hline
    \mathrm{system} & \mathrm{band~SO} & {\bm \Delta}_{\bm k} &
    \sigma^\mathrm{prec}_\lambda & \sigma^\mathrm{sct}_\lambda \\ \hline
    e^- & R1 & {\bm k} & 1/2 & -1/2 \\
    e^- & D1 & {\bm k} & 1/2 & -1/2 \\
    e^- & D3 & {\bm k} & 1/2 & -1/4 \\
    h^+ & R3 & {\bm k}^3 & 0 & 0 \\
    h^+ & D1' & {\bm k}^3 & 0 & 0 \\
    h^+ & D3' & {\bm k}^3 & 0 & 0 \\ \hline\hline
  \end{array}$
\end{table}

\subsection{Anomalous spin precession from electric field}
\label{sec:ASP_E}

Using Eq.\ \ref{TlargeSO}, we have a term in the density matrix
\begin{equation}\label{Ssol1}
  S_{E{\bm k}\lambda}^\mathrm{prec} = - \frac{1}{2} \, \bm{\sigma} \cdot {\bm
    \Delta}_{{\bm k}} \, \frac{\Omega_{\bm k}^2 \, \tau^2}{1 +
    \Omega_{\bm k}^2 \tau^2} \, \delta(\varepsilon_{0{\bm k}} -
  \varepsilon_F).
\end{equation}
In the weak momentum scattering limit $\Omega_{\bm k} \tau \gg 1$
this result is \textit{independent} of the form of the band
structure SO interaction, and can be easily obtained from the driving term in Eq.~(\ref{dsj}).
We have given (in this subsection alone) a
result valid beyond the weak momentum scattering limit so as to
emphasize this apparent independence is only an artifact of this
limit. For electron systems in this limit, the spin-Hall
conductivity due to this term is,
\begin{equation}\label{intsjSHE}
  \begin{array}{rl}
    \sigma^\mathrm{prec}_\lambda = & \displaystyle \frac{n_e e \lambda}{2},
  \end{array}
\end{equation}
where $n_e$ is the electron density. In the weak momentum scattering
limit this term is also independent of $\tau$. In 2D electron
systems it recovers the nonzero contribution to the SHE originally
found by Tse and Das Sarma \cite{TsePRB06} and subsequently by
Raimondi and Schwab.\cite{Raimondi09} In 2D hole systems it is easy
to check that $\sigma^\mathrm{prec}_\lambda = 0$.

The origin of this contribution to the SHE will be elucidated in
Sec.\ \ref{sec:disc}, but one remark is in order here. The spin-Hall
conductivity $\sigma^\mathrm{prec}_\lambda$ found in Eq.\
(\ref{intsjSHE}) has the opposite sign to that found in
Refs.~\onlinecite{TsePRB06, Raimondi09, Culcer_SideJump_PRB10} for
the same orientation of the electric field. One should therefore not
think of $\sigma^\mathrm{prec}_\lambda$ as a \textit{surviving}
side-jump term, but a qualitatively new term due to $\hat{\bm
r}_{so}$ altogether, which we identify with a spin precession
mechanism with no counterpart in systems without band structure SO
coupling.

\subsection{Anomalous spin precession from impurities}

The last piece in the puzzle is the driving term $\hat{J}_{\Omega
\lambda} (n_{E \bm k})$, which needs to be studied independently for
each model. We denote the contribution of this term to
$\sigma_\lambda$ by $\sigma_\lambda^\mathrm{sct}$. Once found, this
term is added to $\sigma_\lambda^\mathrm{prec}$ to give
$\sigma_\lambda$, which yields the total SHE due to $\hat{\bm
r}_{so}$.

\subsubsection{Linear Rashba and Dresselhaus SO}

For linear Rashba band structure SO coupling $H_{R1}$
\begin{equation}
  \begin{array}{rl}
    \displaystyle \hat{J}_{\Omega \lambda} (n_{E \bm k}) = &
    \displaystyle \frac{ 2e \alpha \lambda m k }{\hbar^3 } \,\bm{E}
    \cdot \hat{\bm k} \, {\bm \sigma}\cdot\hat{\bm \theta} \,
    \delta(k - k_F).
  \end{array}
\end{equation}
The spin-Hall conductivity due to this driving term is
\begin{equation}
  \begin{array}{rl}
    \sigma_\lambda^\mathrm{sct} = & \displaystyle - \frac{n_e e
      \lambda}{2}.
  \end{array}
\end{equation}
This term exactly cancels $\sigma_\lambda^\mathrm{prec}$. The same holds
for the linear Dresselhaus SO interaction $H_{D1}$.

\subsubsection{Cubic Dresselhaus SO}
\label{sec:HD3}

In general $\sigma_\lambda^\mathrm{prec}$ and
$\sigma_\lambda^\mathrm{sct}$ do not cancel. We consider next a 2DEG in which the
band structure SO coupling is described by the cubic
Dresselhaus Hamiltonian $H_{D3}$. In this case, the scattering term
$\hat{J}_{\Omega \lambda}^{D3}(n_{E \bm k})$ is given by
\begin{eqnarray}
    \hat{J}_{\Omega \lambda}(n_{E \bm k}) & = &
    - \frac{ m e \beta \lambda k^3 }{ \hbar^3 }
    \,\bm{E} \cdot \hat{\bm k} \,(\bm \sigma\cdot \hat{\bm \theta}
    \sin2\theta - \bm \sigma\cdot \hat{\bm k} \cos2\theta) 
    \nonumber\\ [1ex] & & {} \times \delta(k - k_F).
\end{eqnarray}
This gives a significant  contribution to the spin-Hall current,
\begin{equation}
  \begin{array}{rl}
    \sigma_\lambda^\mathrm{sct} = & -\displaystyle \frac{n_ee\lambda}{4}.
  \end{array}
\end{equation}
The remaining term due to $\hat{\bm r}_{so}$ is $\sigma_\lambda^\mathrm{prec}$, and thus in
the weak momentum scattering limit
\begin{equation}
  \begin{array}{rl}
    \sigma_\lambda = & \displaystyle \frac{n_ee\lambda}{4}.
  \end{array}
\end{equation}
The magnitude of the SHE conductivity due to the band structure SO
coupling (the band-structure SHE) in the 2D cubic Dresselhaus model
has been calculated to be $-e/16\pi$ in the clean limit.
\cite{Malshukov_SHE_2DeDres_PRB05} Therefore the \text{total} SHE
conductivity, including that due to band structure SO, is
\begin{equation}\label{CubicDresSHE}
  \begin{array}{rl}
    \sigma^z_{yx} \approx & \displaystyle - \frac{e}{16\pi} + \frac{n_ee\lambda}{4}.
  \end{array}
\end{equation}
The term due to band-structure SO is density-independent, whereas the anomalous spin precession term in the SHE is linear in $n_e$.
These are the only two terms in the clean limit when the band structure SO coupling is described by the cubic Dresselhaus model.

The cubic Dresselhaus term $H_{D3}$ is strong in a wide quantum well at high electron density $n_e$. However, the full Hamiltonian for such a system in general involves both linear and cubic Dresselhaus SO terms, $H_{D1}$ and $H_{D3}$, whose interplay is nontrivial. We discuss the full conditions required for experimental observation of anomalous spin precession in this complex case in Sec.~\ref{sec:expt}.

\subsubsection{Hole systems}

It is easily seen that in 2D hole systems both
$\sigma_\lambda^\mathrm{prec}$ and $\sigma_\lambda^\mathrm{sct}$ are
zero. For holes, ${\bm \Delta}_{{\bm k}}$ can be found from Eqs.\
(\ref{rsoh}) and (\ref{HE}). Substituting this into Eq.\ \ref{Ssol1},
we find that the spin-Hall current averages to zero over directions
in momentum space. In $\hat{J}_{\Omega \lambda} (n_{E \bm k})$, in
all cases studied, terms $\propto e^{\pm 3i\theta}$ cause the
angular integral to vanish. Therefore, in 2D hole systems
\begin{equation}
  \begin{array}{rl}
    \sigma_\lambda^\mathrm{sct} = & \displaystyle 0,
  \end{array}
\end{equation}
There is thus no contribution to the SHE due to anomalous spin
precession in 2D hole systems.

\section{Discussion}
\label{sec:disc}

To summarize, $\sigma_\lambda = 0$ in 2D hole systems, while in 2D
electron systems in the weak momentum scattering regime it can be
written as
\begin{equation}
  \begin{array}{rl}
    \sigma_\lambda = & \displaystyle \frac{n_ee\lambda}{2} +
    \sigma_\lambda^{sct}.
  \end{array}
\end{equation}
The results for the total SHE due to $\hat{\bm r}_{so}$ are
summarized in Table \ref{tab:disentangle}. Interestingly,
$\sigma_\lambda$ can be nonzero, even though that is only true in
one out of the several situations studied explicitly in this work.

We have argued previously that $\sigma_\lambda^\mathrm{prec}$ should
be thought of not as a surviving side-jump term, but a qualitatively
new term, which is not present in systems without band structure SO.
We demonstrate that this term is related to spin precession induced
by both band structure SO and $\hat{\bm r}_{so}$. The electric field
${\bm E}$ gives rise to an additional SO effective field ${\bm
\Delta}_{\bm k} \parallel \hat{\bm z}$-direction. The band structure
SO effective field ${\bm \Omega}_{\bm k}$ is in the plane. We
examine spin precession in the total effective magnetic field ${\bm
\Omega}_{\bm k}$ and ${\bm \Delta}_{\bm k}$, redefining ${\bm
\Omega}_{\bm k} \ra \tilde{\bm \Omega}_{\bm k}$, with
\begin{equation}
  \tilde{{\bm \Omega}}_{\bm k} = {\bm \Omega}_{\bm k} + {\bm
    \Delta}_{\bm k}.
\end{equation}
Let $\tilde{{\bm \Omega}}_{\bm k} = (\tilde{\Omega}_x, 0, 0)$ and
turn on ${\bm E}$ adiabatically, generating a small
$\tilde{\Omega}_z \ll \tilde{\Omega}_x$. We study the Heisenberg
equation of motion for the spin (Bloch) vector ${\bm s}$, which
reads $d{\bm s}/dt = \tilde{{\bm \Omega}} \times {\bm s}$, in a
clean system. The spin is taken initially to be parallel to
$\tilde{\Omega}_x$. In component form
\begin{subequations}
  \begin{eqnarray}
    \displaystyle \td{s_x}{t} & = & \displaystyle - \tilde{\Omega}_z
    s_y, \\ [2ex]
    \label{sy}
    \displaystyle \td{s_y}{t} & = & \displaystyle \tilde{\Omega}_z
    s_x - \tilde{\Omega}_x s_z, \\ [2ex] \displaystyle \td{s_z}{t} &
    = & \displaystyle \tilde{\Omega}_x s_y.
  \end{eqnarray}
\end{subequations}
One can take the time derivative one more time and solve the
equations exactly, yet the physics is evident from Eq.\ (\ref{sy}).
Since ${\bm s} (t = 0) = (s_x, 0, 0)$ and $s_y$ is initially zero,
$s_y$ should remain zero at all times. Setting $ds_y/dt$ in the
steady state we obtain
\begin{equation}
  s_z = s_x \,
  \bigg(\frac{\tilde{\Omega}_z}{\tilde{\Omega}_x}\bigg).
\end{equation}
The explanation is as follows: $s_y$ is initially 0 and must remain
0. When ${\bm E}$ is turned on an additional component
$\tilde{\Omega}_z$ is generated, which makes $s_x$ precess and gives
a small contribution to $s_y$. To cancel this, $s_z$ must develop a
small out-of-plane component, which precesses around
$\tilde{\Omega}_x$, and gives the exact opposite contribution to
$s_y$. The extra $s_z$ density has opposite signs for the two halves
of the Fermi surface, giving rise to a net spin-Hall current. The
argument presented here shows that $\hat{\bm r}_{so}$ gives rise to
a spin-Hall current even in a clean system. We refer to this process
as \textit{anomalous spin precession}.

This argument can be generalized to explain anomalous spin
precession in a disordered system as well. This can be done by
replacing ${\bm E} \rightarrow {\bm E} + {\bm \nabla} U({\bm r})$,
and understanding this to represent the total \textit{local}
electric field. We thus reproduce both anomalous spin precession
terms -- the one due to the external electric field and the one due
to the impurity potential. Both terms give an effective magnetic
field out of the plane of the quantum well, modifying the intrinsic
SO spin precession.

Equation (\ref{TlargeSO}) is valid for weak momentum scattering.
Appendix \ref{sec:ST} shows that in the strong momentum scattering
regime $\overline{S_{E{\bm k}\lambda}}$ diverges. Physically, this
is because we are using ${\bm \Omega}_{\bm k}$ as our reference, and
projections parallel and perpendicular to it become ill-defined as
${\bm \Omega}_{\bm k} \rightarrow 0$. In this limit Dyakonov-Perel
spin relaxation is no longer active, and there is no spin relaxation
at all. We demonstrate in Appendix \ref{sec:ST} that the divergence
in the strong momentum scattering regime is cured by the
introduction of the Elliott-Yafet spin relaxation time
$\tau_\mathrm{EY}$, which is also related to $\mathcal{V}_{{\bm
k}{\bm k}'}$. Nevertheless, in order to be consistent one would have
to formulate the entire theory up to order $\lambda^2$, which is
beyond the scope of this paper.

In deriving $\sigma_\lambda$ we have assumed for simplicity that the
scattering potential is short ranged. We do not expect the results
to change qualitatively for long-range impurities. Firstly, we have
shown that $\sigma_\lambda^\mathrm{prec}$ is independent of
scattering in weak momentum scattering limit and is traced to a
mechanism unrelated to disorder. Secondly, although for a
general potential the anisotropic terms in $\mathcal{U}_{{\bm k}{\bm
k}'}$ will depend on the form of the potential, as will
$\sigma^\mathrm{sct}_\lambda$, we do not expect cancellation between
$\sigma^\mathrm{sct}_\lambda$ and $\sigma^\mathrm{prec}_\lambda$,
even though $\sigma^\mathrm{sct}_\lambda$ may have a different
numerical value from that determined.  Finally, past experience
with the SHE shows that important cancellations, such as that of the
SHE due to Rashba band structure SO coupling, tend to have a
fundamental origin \cite{Dimitrova05} and are independent of whether
the scattering potential is short-range or long-range.
\cite{Culcer_SteadyState_PRB07, Culcer_Generation_PRL07}

\section{Experimental observation}
\label{sec:expt}

We have argued that the anomalous spin precession contribution to the SHE in general is finite. For example it survives in 2D electron gases in which the SO interaction is described by the cubic Dresselhaus term ($H_{D3}$). In Sec.~\ref{sec:HD3} we calculated the anomalous spin precession contribution to the SHE conductivity using purely the cubic Dresselhaus model. We now discuss the experimental conditions required for the observation of anomalous spin precession in a realistic sample.

For the anomalous spin precession contribution to the SHE to be observable it must ideally overwhelm the band structure contribution. Here we focus on two common semiconductor materials with strong SO coupling in the conduction band, InAs and InSb, and estimate the magnitude of the anomalous spin precession as well as the band structure contributions to the SHE in these materials. The constant $\lambda$ for InAs and InSb can be found in Table 6.6
in Ref.\ \onlinecite{Winkler2003} (in the notation used in this
paper, $\lambda = r^{6c6c}_{41}/e$). 

The situation is complicated by the fact
that in realistic 2D samples both the linear and the cubic
Dresselhaus terms, $H_{D1}$ and $H_{D3}$, are present. Having noted in Sec.\ \ref{sec:BandHam}
that $\beta_1 \simeq \beta_3 (\pi/w)^2$, the total SO Hamiltonian
is
\begin{equation}
  \arraycolsep 0.3ex
  \begin{array}[b]{rl}
    H = \displaystyle \frac{\beta_3 \pi^2}{w^2} \, (\sigma_y k_y - \sigma_x k_x) + \beta_3 (\sigma_x k_x k_y^2 - \sigma_y k_y k_x^2) 
  \end{array}
\end{equation}
The ratio $\pi/(k_Fw)$ determines the relative magnitudes of $H_{D1}$ and $H_{D3}$. However, in order to have only one subband
occupied it is necessary that $\pi/(k_Fw) \ge 1$. 

\begin{figure}[tbp]
\includegraphics[width=\columnwidth]{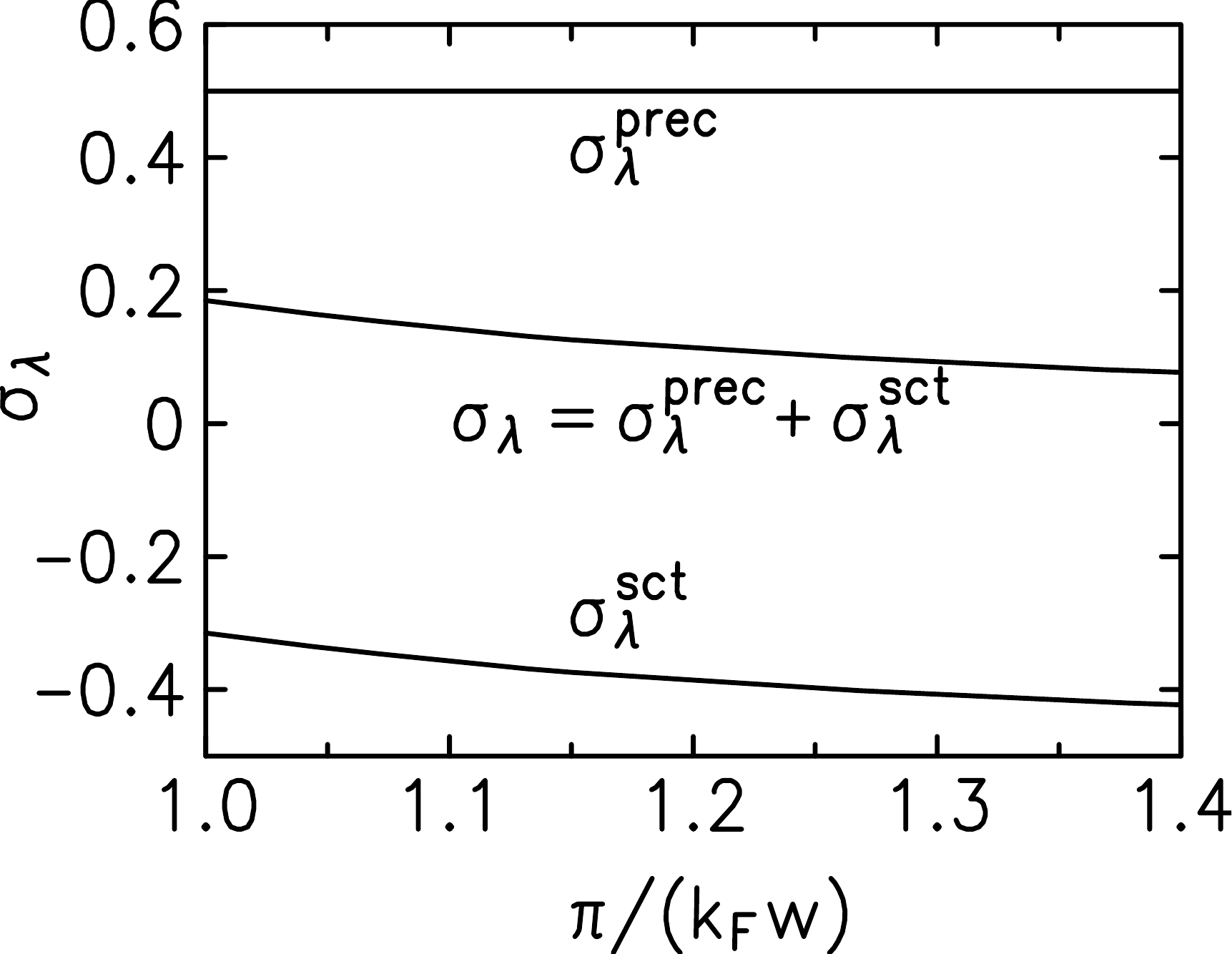}
\caption{\label{plot} Anomalous spin precession contributions in a 2D electron gas with band structure spin-orbit coupling described by $H = H_{D1} + H_{D3}$ as a function of the parameter $\pi/(k_F w)$. On the vertical axis $\sigma_\lambda$ is measured in units of $n_ee\lambda$.
}
\end{figure}

We showed in Sec.\ \ref{sec:ASP_E} that $\sigma_\lambda^{prec}$ is the same in the clean limit independently of the form of the band structure spin-orbit coupling. On the other hand, the contributions of the linear and cubic Dresselhaus terms, $H_{D1}$ and $H_{D3}$, to $\sigma_\lambda^{sct}$ are not simply additive, and their interplay is nontrivial. Therefore, the calculation of $\sigma_\lambda^{sct}$ presented in Sec.~\ref{sec:HD3} needs to be repeated for the complicated case of $H = H_{D1} + H_{D3}$. This is done here analytically, except that the final results require a series of lengthy numerical integrations which can be performed using a symbolic algebra package. The results for $\sigma_\lambda^{prec}$ and $\sigma_\lambda^{sct}$ are summarized in Table~\ref{tab:HD1HD3}, as well as Fig.~\ref{plot}.

The band structure contribution to the SHE for the case $H = H_{D1} + H_{D3}$ has been evaluated in Ref.\
\onlinecite{Malshukov_SHE_2DeDres_PRB05}. In Fig.~1 of that reference it was shown that the band structure SHE is a
non-monotonic function of the parameter $\pi/(k_Fw)$, where $\pi/w$
in our paper corresponds to the parameter $a$ in Ref.\
\onlinecite{Malshukov_SHE_2DeDres_PRB05}. In fact, the band structure SHE conductivity varies
strongly as a function of this parameter and it changes sign at a
critical value. It is however \textit{independent} of $\beta_3$ in the clean limit, as is customary in 2D electron gases.

We consider a high-mobility quantum well with a number density $n_e = 5 \times
10^{12}$~cm$^{-2}$ for concreteness, a density commonly encountered
in transport experiments. We focus on values of $w$ for which $\pi/(k_Fw) $ is comprised
between $1.0$ (the widest well) and $1.4$.

The band structure contribution including \textit{both linear and cubic
terms} is read off from Fig.~1 of Ref.\
\onlinecite{Malshukov_SHE_2DeDres_PRB05} and is the same for InAs
and InSb. Our Eq.\ (\ref{CubicDresSHE}) (the pure cubic case)
corresponds to $a=0$ in Eq. (16) of Ref.\
\onlinecite{Malshukov_SHE_2DeDres_PRB05}. Note also that, in the
notation of Ref.\ \onlinecite{Malshukov_SHE_2DeDres_PRB05}, $e$
denotes the electron charge, whereas in our notation the electron
charge is $-e$: hence the seemingly opposite sign of the first term
of our Eq.\ (\ref{CubicDresSHE}) compared to the corresponding
formula of Ref.\ \onlinecite{Malshukov_SHE_2DeDres_PRB05}. When
$\pi/(k_Fw) = 1.0$, the band structure contribution is $\approx 0.8
\times e/(16\pi) \approx 0.016~e$ and, referring to
Table~\ref{tab:HD1HD3}, we find the anomalous spin precession
contribution to be $\approx 0.185~n_ee\lambda$. When $\pi/(k_Fw) =
1.4$, the band structure contribution decreases to $\approx 0.2
\times e/(16\pi) \approx 0.004~e$, and the anomalous spin precession
contribution to $\approx 0.077~n_ee\lambda$.

We consider first InAs, for which $\lambda = 117$~\AA$^2$. At
$\pi/(k_Fw) = 1.0$, with the value of $n_e$ specified above, we
find the anomalous spin precession term to be $\approx 0.01~e$, which
is just over half the size of the band structure term. At
$\pi/(k_Fw) = 1.4$, the anomalous spin precession term is $0.0045~e$,
marginally larger than the band structure term. In InAs therefore
the band structure term is dominant in this parameter range.

In InSb, on the other hand, $\lambda = 523$~\AA$^2$. At $\pi/(k_Fw) =
1.0$, with the value of $n_e$ given above, we find the anomalous spin
precession term to be $\approx 0.05~e$, three times larger than the
band structure term. At $\pi/(k_Fw) = 1.4$, the anomalous spin
precession term is $0.02~e$, five times larger than the band
structure term. Thus, in InSb the anomalous spin precession is
dominant in this parameter range.

\begin{table}[tbp]
  \caption{\label{tab:HD1HD3} Anomalous spin precession
  contributions to the spin-Hall conductivity in a 2D electron gas
  in a cubic crystal, with band-structure spin-orbit described by $H = H_{D1} + H_{D3}$, all in units of $n_ee\lambda$. In the last
  column $\sigma_\lambda = \sigma_\lambda^{prec} +
  \sigma_\lambda^{sct}$.}
  $\arraycolsep 1em
   \begin{array}{c@{\hspace{2em}}cccc} \hline\hline
  \pi/(k_Fw) & \sigma_\lambda^{prec} & \sigma_\lambda^{sct} & \sigma_\lambda 
  \\  \\ \hline
 1.00 & 0.5 & -0.315 & 0.185 \cr   
 1.05 & 0.5 & -0.338 & 0.162 \cr 
 1.10 & 0.5 & -0.357 & 0.143 \cr 
 1.15 & 0.5 & -0.374 & 0.126 \cr    
 1.20 & 0.5 & -0.386 & 0.114 \cr 
 1.25 & 0.5 & -0.398 & 0.102 \cr 
 1.30 & 0.5 & -0.407 & 0.093 \cr 
 1.35 & 0.5 & -0.416 & 0.084 \cr 
 1.40 & 0.5 & -0.423 & 0.077  
 \\ \hline \hline
  \end{array}$
\end{table}

We conclude that the most promising system for the observation of
anomalous spin precession is the 2D electron gas in InSb. In the
range $ 1.0 \le \pi/(k_Fw) \le 1.4$ the anomalous spin precession
provides the dominant contribution to the spin-Hall effect. At the
lower end of this range, the overall SHE signal is stronger, and
anomalous spin precession accounts for approximately three quarters
of the SHE conductivity. At the upper end, although the overall
signal is weaker, anomalous spin precession accounts for
approximately $5/6$ of the SHE conductivity.

\section{Summary and conclusions}

We have determined all the contributions to the SHE due to the
anomalous position operator $\hat{\bm r}_{so}$ in 2D electron and
hole systems. The SHE due to skew scattering and side-jump
scattering vanishes in the presence of spin precession caused by the
band structure SO coupling. Two additional contributions to the SHE
exist due to $\hat{\bm r}_{so}$, one of which is
scattering-dependent and one of which is due to \textit{anomalous}
spin precession under the action of $\hat{\bm r}_{so}$ and the
electric field. These two contributions cancel out in systems with
band structure SO linear in ${\bm k}$, and are independently zero in
2D hole systems. However, the contribution due to anomalous spin
precession survives in 2D electron systems with a significant cubic
Dresselhaus term, i.e., for wide quantum wells with high electron
densities, and is dominant under certain circumstances in InSb.
Anomalous spin precession can therefore be detected in such a
system.

A full account of the SHE in 2D systems must include the lengthy
calculation of the electric field contribution to the skew
scattering term, plus the band structure SO correction to that term.
Moreover, in this work we have only considered heterostructures
grown along the main crystal axes. Finally, the full answer will be
known when the definition of the conserved spin current is taken
into account, as has been done for the band-structure SHE.
\cite{Sugimoto_SHE_CnsvdCrnt_PRB06} We reserve these studies for a
future publication.

\acknowledgments

We acknowledge insightful discussions with S.~Das Sarma,
O.~P.~Sushkov, Peter Schwab, Roberto Raimondi, Cosimo Gorini and
W.~K.~Tse. DC was in part supported by the Chinese Academy of
Sciences. E.~M.~H.\ was financially supported by the German Science
Foundation DFG grant HA 5893/1-2 within the SPP 128. GV acknowledges
support from NSF Grant No. DMR-1104788.
Work at Argonne was supported by DOE BES under Contract
No.\ DE-AC02-06CH11357.

\appendix

\begin{widetext}

\section{Decomposition of the spin density matrix into
  $\overline{S_{E{\bm k}\lambda}}$ and $T_{E{\bm k}\lambda}$}
\label{sec:ST}

>From the quantum Liouville equation, we obtain for
$\overline{S_{E{\bm k}\lambda}}$ and $T_{E{\bm k}\lambda}$ for
short-range impurities
\begin{subequations}\label{SbarT}
  \begin{eqnarray}
    \label{KineqS}
    \displaystyle \pd{ \overline{S_{E{\bm k}\lambda}} }{t} +
    \frac{i}{\hbar} \, \overline{[H, T_{E{\bm k}\lambda}]} & = &
    \overline{\mathcal{D}_{E{\bm k}\lambda}} \\ [1ex]
    \label{KineqT}
    \displaystyle \pd{T_{E{\bm k}\lambda}}{t} + \frac{i}{\hbar} \,
                  {[H, T_{E{\bm k}\lambda}]} + \frac{T_{E{\bm
                        k}\lambda}}{\tau} \displaystyle & = &
                  (\mathcal{D}_{E{\bm k}\lambda} -
                  \overline{\mathcal{D}_{E{\bm k}\lambda}}) -
                  \frac{i}{\hbar} \, {[H, \overline{S_{E{\bm
                            k}\lambda}}]} + \frac{i}{\hbar} \,
                  \overline{{[H, T_{E{\bm k}\lambda}]}}.
  \end{eqnarray}
\end{subequations}
On the RHS of Eq.\ (\ref{KineqT}) we substitute for $\displaystyle
\frac{i}{\hbar} \, \overline{{[H, T_{E{\bm k}\lambda}]}}$ from
Eq.\ (\ref{KineqS}). We rewrite Eqs.\ (\ref{SbarT}) as
\begin{subequations}
  \begin{eqnarray}
    \label{eq:rhobarga}
    \displaystyle \pd{\overline{S_{E{\bm k}\lambda}}}{t} +
    \frac{i}{\hbar} \, \overline{{[H_{\bm k}, T_{E{\bm k}\lambda}]}}
    & = & \overline{\mathcal{D}_{E{\bm k}\lambda}} \\ [1ex]
    \label{eq:rhobargb}
    \displaystyle \pd{T_{E{\bm k}\lambda}}{t} + \frac{i}{\hbar} \,
                  {[H_{\bm k}, T_{E{\bm k}\lambda}]} +
                  \frac{T_{E{\bm k}\lambda}}{\tau} \displaystyle & =
                  & \mathcal{D}_{E{\bm k}\lambda} -
                  \bigg(\pd{\overline{S_{E{\bm k}\lambda}}}{t} +
                  \frac{i}{\hbar} \, [H_{\bm k}, \overline{S_{E{\bm
                          k}\lambda}}] \bigg).
  \end{eqnarray}
\end{subequations}
Defining $T_{E{\bm k}\lambda} = e^{- i H_{\bm k} t/\hbar}
\tilde{T}_{E{\bm k}\lambda} \, e^{i H_{\bm k} t/\hbar}$ and
$\overline{S_{E{\bm k}\lambda}} = e^{- i H_{\bm k} t/\hbar}
\tilde{\overline{S_{E{\bm k}\lambda}}} \, e^{i H_{\bm k} t/\hbar}$,
we can easily solve Eq.\ (\ref{eq:rhobargb})
\begin{subequations}
  \begin{eqnarray}
    \displaystyle \pd{\tilde{T}_{E{\bm k}\lambda}}{t} +
    \frac{\tilde{T}_{E{\bm k}\lambda}}{\tau} & = & \displaystyle
    e^{i H_{\bm k} t/\hbar} \mathcal{D}_{E{\bm k}\lambda} e^{- i
    H_{\bm k} t/\hbar} - \pd{\tilde{\overline{S_{E{\bm k}\lambda}}}}
    {t} \\ \displaystyle \tilde{T}_{E{\bm k}\lambda} & = &
    \displaystyle - \tilde{\overline{S_{E{\bm k}\lambda}}} +
    \displaystyle \int_{-\infty}^{t} dt' \, e^{-\frac{(t -
    t')}{\tau}} \bigg[e^{i H t'/\hbar} \mathcal{D}_{E{\bm k}\lambda}
    e^{- i H t'/\hbar} + \frac{\tilde{\overline{S_{E{\bm
    k}\lambda}}}}{\tau} \bigg].
  \end{eqnarray}
\end{subequations}
where the last line was obtained by integration by parts. We can
write $T_{E{\bm k}\lambda}$ (without the tilde) as
\begin{equation}\label{Tsol1}
  \begin{array}{rl}
    \displaystyle T_{E{\bm k}\lambda} = & \displaystyle -
    \overline{S_{E{\bm k}\lambda}} + \int_0^{\infty} dt' \,
    e^{-\frac{t'}{\tau}} e^{-i H t'/\hbar} \bigg(\mathcal{D}_{E{\bm
        k}\lambda} + \frac{\overline{S_{E{\bm k}\lambda}}}{\tau}
    \bigg) e^{i H t'/\hbar}.
  \end{array}
\end{equation}
Using $\displaystyle \overline{S_{E{\bm k}\lambda}} = \frac{1}{2} \,
{\bm \sigma} \cdot \overline{{\bm s}_{E{\bm k}\lambda}}$,
$\displaystyle \mathcal{D}_{E{\bm k}\lambda} = \frac{1}{2} \, {\bm
  \sigma} \cdot {\bm d}_{E{\bm k}\lambda}$ and $\displaystyle
T_{E{\bm k}\lambda} = \frac{1}{2} \, {\bm \sigma} \cdot {\bm
  t}_{E{\bm k}\lambda}$, and carrying out the time integral
\begin{equation}\label{Tsol2}
  \begin{array}{rl}
    \displaystyle {\bm t}_{E{\bm k}\lambda} = & \displaystyle
    \hat{\bm \Omega}_{\bm k} \times \bigg({\bm d}_{E{\bm k}\lambda}
    + \frac{\overline{\bm s}_{E{\bm k}\lambda}}{\tau} \bigg) \,
    \frac{\Omega_{\bm k} \tau^2}{1 + \Omega_{\bm k}^2 \tau^2} +
    \frac{ ({\bm d}_{E{\bm k}\lambda}\tau)}{1 + \Omega_{\bm
        k}^2\tau^2} + additional \,\,\, terms.
  \end{array}
\end{equation}
The physical interpretation of the terms appearing in Eq.\
\ref{Tsol2} is as follows. The first term [containing $\hat{\bm
\Omega}_{\bm k} \times (\ldots)$] gives the full spin current when
there is spin precession ($\Omega_{\bm k} \ne 0$). The second term
(containing ${\bm d}_{E{\bm k}\lambda}\tau$) recovers the spin
current due to \textit{impurity} SO coupling when there is no spin
precession ($\Omega_{\bm k} = 0$). It vanishes in the weak momentum
scattering limit $\Omega_{\bm k} \tau \gg 1$. Finally, the
\textit{additional terms} ensure that ${\bm t}_{E{\bm k}\lambda}$
averages to zero over directions in momentum space, but these terms
give no spin current.

Let $\mathcal{A}_{ij} = (\delta_{ij} - \hat{\Omega}_{{\bm
    k}i}\hat{\Omega}_{{\bm k}j})$, abbreviate
$\mathcal{A}\overline{\bm s}_{E{\bm k}\lambda} \equiv
\mathcal{A}_{ij} \overline{s}_{{E{\bm k}\lambda}, j}$, and substitue
Eq.\ (\ref{Tsol2}) into Eq.(\ref{eq:rhobarga}). In the steady state
\begin{equation}\label{Ssol}
  \begin{array}{rl}
    \displaystyle \frac{1}{\tau} \, \bigg[
      \bigg(\overline{\frac{\Omega_{\bm k}^2\tau^2}{1 + \Omega_{\bm
            k}^2\tau^2}\bigg) \mathcal{A}} \bigg] \, \overline{{\bm
        s}_{E{\bm k}\lambda}} = & \displaystyle \overline{{\bm
        d}_{E{\bm k}\lambda}} - \bigg(\overline{\frac{\Omega_{\bm
          k}^2\tau^2}{1 + \Omega_{\bm k}^2\tau^2} \bigg)
      \mathcal{A}{\bm d}_{E{\bm k}\lambda}} + \overline{ \frac{({\bm
          \Omega_{\bm k}} \times {\bm d}_{E{\bm k}\lambda}) \,
        \tau}{1 + \Omega_{\bm k}^2 \tau^2}},
  \end{array}
\end{equation}
using $\displaystyle \frac{i}{\hbar} \, [H_{\bm k},
  \overline{S_{E{\bm k}\lambda}}] = - \frac{1}{2} \, {\bm \sigma}
\cdot {\bm \Omega_{\bm k}} \times \overline{\bm s}_{E{\bm
    k}\lambda}$. For $\Omega_{\bm k}\tau \gg 1$ we obtain simply
\begin{subequations}
  \begin{eqnarray}\label{STlargeSO}
    \displaystyle
    \bigg(\frac{\overline{\mathcal{A}}}{\tau}\bigg) \,
    \overline{{\bm s}_{E{\bm k}\lambda}} & = & \displaystyle
    \overline{{\bm d}_{E{\bm k}\lambda}} -
    \overline{\mathcal{A}{\bm d}_{E{\bm k}\lambda}} \\
    \displaystyle {\bm t}_{E{\bm k}\lambda} & = & \displaystyle
    \bigg(\frac{\hat{\bm \Omega}_{\bm k}}{\Omega_{\bm k}}\bigg)
    \times \bigg({\bm d}_{E{\bm k}\lambda} + \frac{\overline{\bm
    s}_{E{\bm k}\lambda}}{\tau} \bigg) + additional \,\,\,
    terms.
  \end{eqnarray}
\end{subequations}
\section{Elliott-Yafet spin relaxation time}

This derivation is for a general $S_{\bm k}$. Consider the
scattering term in the Born approximation Eq.\ (\ref{JBorn}) up to
\textit{second} order in $\lambda$, and focus on its action on
$S_{\bm k}$. In this term we may ignore the part of the time
evolution operator $\propto \Omega_{\bm k}$. This scattering term is
referred to as $\hat{J}_\mathrm{EY} (S_{\bm k})$, and takes the form
\begin{subequations}
  \begin{eqnarray}
    \displaystyle \hat{J}_\mathrm{EY} (S_{\bm k}) & = &
    \frac{\pi n_i}{\hbar} \int \frac{d^2k'}{(2\pi)^2} \,
    \mathcal{V}_{\bm{k}\bm{k}'} \, (
    \mathcal{V}_{\bm{k}'\bm{k}}S_{\bm k} - S_{\bm
    k'}\mathcal{V}_{\bm{k}'\bm{k}} ) \, \delta(\varepsilon_{\bm k} -
    \varepsilon _{{\bm k}'} ) + h.c. \\ [1ex] & = & 
    \frac{n_im}{2\hbar^3} \int_0^{2\pi} \frac{d\theta'}{2\pi} \, (
    |\mathcal{V}_{{\bm k} {\bm k}'}|^2S_{\bm k} -
    \mathcal{V}_{\bm{k}\bm{k}'} \, S_{\bm
    k'}\mathcal{V}_{\bm{k}'\bm{k}}) + h.c.
  \end{eqnarray}
\end{subequations}
Bearing in mind that $|\mathcal{V}_{{\bm k} {\bm k}'}|^2$ is a
scalar, and in 2D systems $\mathcal{V}_{{\bm k} {\bm k}'} \propto
\sigma_z$, the term $\mathcal{V}_{\bm{k}\bm{k}'} \, S_{\bm
  k'}\mathcal{V}_{\bm{k}'\bm{k}}$ has two possible forms:
\begin{equation}
    \displaystyle \mathcal{V}_{\bm{k}\bm{k}'} \, S_{\bm k'}
    \mathcal{V}_{\bm{k}'\bm{k}} = 
    \left\{ \begin{array}{ll} |\mathcal{V}_{{\bm k} {\bm k}'}|^2 \,
      S_{\bm k'}, \,\,\, & {\rm for} \,\,\, S_{\bm k'} \propto
      \sigma_z \\[1ex] - |\mathcal{V}_{{\bm k} {\bm k}'}|^2 \, S_{\bm
        k'}, \,\,\, & {\rm for} \,\,\, S_{\bm k'} \propto \sigma_x,
      \sigma_y
    \end{array}\right.
\end{equation}
so that
\begin{equation}
  \hat{J}_\mathrm{EY} (S_{\bm k}) =
    \displaystyle \frac{n_im}{\hbar^3} \int_0^{2\pi}
    \frac{d\theta'}{2\pi} \, |\mathcal{V}_{{\bm k} {\bm
        k}'}|^2(S_{\bm k} - m_z S_{\bm k'}),
\end{equation}
where $m_z = -1$ before $\sigma_x, \sigma_y$ and $m_z = 1$ before
$\sigma_z$. If $S_{\bm k} \propto \sigma_z$ the spin is out of the
plane and is conserved during scattering, thus $\hat{J}_\mathrm{EY} (S_{\bm
  k})$ gives just a correction to the momentum relaxation time. The
change of sign for $S_{\bm k} \propto \sigma_x, \sigma_y$ is
crucial. For short-range impurities, with $|\mathcal{V}_{{\bm k}
  {\bm k}'}|^2 = \lambda^2 k^4 |\mathcal{U}|^2\sin^2\gamma$,
\begin{equation}
  \begin{array}{rl}
    \displaystyle \hat{J}_\mathrm{EY} (S_{\bm k}) = & \displaystyle
    \frac{\lambda^2 k^4}{2\tau} \int_0^{2\pi} \frac{d\theta'}{2\pi}
    \, (S_{\bm k} - m_z S_{\bm k'}) \, (1 - \cos 2\gamma).
  \end{array}
\end{equation}
If we now write $S_{\bm k} = \overline{S_{\bm k}} + T_{\bm k}$, and
define $(1/\tau_\mathrm{EY}) = \lambda^2 k^4/\tau$, then $\hat{J}_\mathrm{EY}
(S_{\bm k})$ simplifies to
\begin{equation}\label{JEY}
  \begin{array}{rl}
    \displaystyle \hat{J}_\mathrm{EY} (S_{\bm k}) = & \displaystyle
    \frac{S_{\bm k} - m_z \overline{S_{\bm k}}}{2\tau_\mathrm{EY}} +
    \frac{m_z}{2\tau_\mathrm{EY}}\int_0^{2\pi} \frac{d\theta'}{2\pi} \,
    T_{\bm k'} \, \cos 2\gamma.
  \end{array}
\end{equation}

\section{$\tau_\mathrm{EY}$ cures divergence in $\overline{S_{E{\bm k}\lambda}}$}

Equations (\ref{STlargeSO}) are correct as long as $\Omega\tau \gg
1$, otherwise $\overline{{\bm s}_{E{\bm k}\lambda}}$ found from Eq.\
(\ref{Ssol}) diverges at small $\Omega\tau$. The way out of this
dilemma is provided by the Elliott-Yafet spin relaxation time.
Consider adding $\hat{J}_\mathrm{EY} (S_{\bm k})$ to Eqs.\
(\ref{SbarT})
\begin{subequations}\label{SbarEY}
  \begin{eqnarray}
    \label{KineqSEY}
    \displaystyle \pd{\overline{S_{E{\bm k}\lambda}}}{t} +
    \frac{i}{\hbar} \, \overline{{[H, T_{E{\bm k}\lambda}]}} +
    \frac{S_{E{\bm k}\lambda} - m_z \overline{S_{E{\bm
            k}\lambda}}}{2\tau_\mathrm{EY}} & = &
    \overline{\mathcal{D}_{\bm k}} \\ [1ex]
    \label{KineqTEY}
    \displaystyle \pd{T_{E{\bm k}\lambda}}{t} + \frac{i}{\hbar} \,
                  {[H, T_{E{\bm k}\lambda}]} + \frac{T_{E{\bm
                        k}\lambda}}{\tau_{tot}} + \hat{J}_\mathrm{EY}
                  (T_{E{\bm k}\lambda}) \displaystyle & = &
                  (\mathcal{D}_{\bm k} - \overline{\mathcal{D}_{\bm
                      k}}) - \frac{i}{\hbar} \, {[H,
                      \overline{S_{E{\bm k}\lambda}}]} +
                  \frac{i}{\hbar} \, \overline{{[H, T_{E{\bm
                            k}\lambda}]}},
  \end{eqnarray}
\end{subequations}
where $1/\tau_{tot} = 1/\tau + 1/\tau_\mathrm{EY}$. Since $\lambda k_F^2
\ll 1$, the Elliott-Yafet spin relaxation time $\tau_\mathrm{EY} \gg \tau$,
and the term containing the angular integral over $\theta'$ is a
very small correction to Eq.\ (\ref{KineqTEY}), which may be
neglected. The only change to the above formalism is an extra term
in the equation for $\overline{S_{E{\bm k}\lambda}}$, which is
nonzero for $\overline{S_{E{\bm k}\lambda}}$ in plane. The spin
generated by an electric field is in-plane, so we can focus on this
component, for which $m_z = -1$, and Eq.\ (\ref{Ssol}) becomes
\begin{equation}\label{SsolEY}
  \begin{array}{rl}
    \displaystyle \overline{\bigg(\frac{\Omega_{\bm
          k}^2\tau_{tot}^2}{1 + \Omega_{\bm k}^2\tau_{tot}^2}\bigg)
      \, \mathcal{A}} \, \frac{\overline{{\bm s}_{E{\bm
            k}\lambda}}}{\tau_{tot}} + \frac{\overline{{\bm
          s}_{E{\bm k}\lambda}}}{\tau_\mathrm{EY}} = & \displaystyle
    \overline{{\bm d}_{E{\bm k}\lambda}} - \overline{\mathcal{A} \,
      {\bm d}_{E{\bm k}\lambda} \, \bigg(\frac{\Omega_{\bm
          k}^2\tau_{tot}^2}{1 + \Omega_{\bm
          k}^2\tau_{tot}^2}\bigg)}.
  \end{array}
\end{equation}
This cures the unphysical divergence at small $\Omega_{\bm k}\tau$.
To see this, consider the simplest case, that of isotropic
$\Omega_{\bm k}$,
\begin{equation}\label{}
  \begin{array}{rl}
    \displaystyle \overline{{\bm s}_{E{\bm k}\lambda}} = &
    \displaystyle \frac{2\overline{{\bm d}_{E{\bm k}\lambda}}
      \tau_{tot} (1 + \Omega_{\bm k}^2\tau_{tot}^2) -
      2\overline{\mathcal{A} \, {\bm d}_{E{\bm k}\lambda}}
      \,\Omega_{\bm k}^2\tau_{tot}^3}{[\Omega_{\bm k}^2\tau_{tot}^2
        + (2\tau_{tot}/\tau_\mathrm{EY}) (1 + \Omega_{\bm
          k}^2\tau_{tot}^2)]}.
  \end{array}
\end{equation}
Clearly $\overline{{\bm s}_{E{\bm k}\lambda}} \rightarrow 0$ as
$\Omega_{\bm k} \rightarrow 0$. Physically, the Elliott-Yafet spin
relaxation time is needed to cure this divergence because
projections parallel and perpendicular to ${\bm \Omega}_{\bm k}$ are
ill-defined as ${\bm \Omega}_{\bm k} \rightarrow 0$.
\end{widetext}


\end{document}